
\input harvmac
%
\noblackbox



\newcount\figno
\figno=0
\def\IZ{\relax\ifmmode\mathchoice
{\hbox{\cmss Z\kern-.4em Z}}{\hbox{\cmss Z\kern-.4em Z}}
{\lower.9pt\hbox{\cmsss Z\kern-.4em Z}} {\lower1.2pt\hbox{\cmsss
Z\kern-.4em Z}}\else{\cmss Z\kern-.4em Z}\fi}
\def\IR{\relax{\rm I\kern-.18em R}}
\def\fig#1#2#3{
\par\begingroup\parindent=0pt\leftskip=1cm\rightskip=1cm\parindent=0pt
\baselineskip=11pt \global\advance\figno by 1 \midinsert
\epsfxsize=#3 \centerline{\epsfbox{#2}} \vskip 12pt
\centerline{{\bf Figure \the\figno} #1}\par
\endinsert\endgroup\par}
\def\figlabel#1{\xdef#1{\the\figno}}

\def\pmb#1{\setbox0=\hbox{#1}%
\kern-.025em\copy0\kern-\wd0 \kern.05em\copy0\kern-\wd0
\kern-.025em\raise.0433em\box0 } \font\cmss=cmss10
\font\cmsss=cmss10 at 7pt
\def\half{{1\over 2}}
\def\rlx{\relax\leavevmode}
\def\Cop{\relax\,\hbox{$\kern-.3em{\rm C}$}}
\def\Rop{\relax{\rm I\kern-.18em R}}
\def\Nop{\relax{\rm I\kern-.18em N}}
\def\Pop{\relax{\rm I\kern-.18em P}}

\def\Zop{\rlx\leavevmode\ifmmode\mathchoice{\hbox{\cmss Z\kern-.4em
Z}} {\hbox{\cmss Z\kern-.4em Z}}{\lower.9pt\hbox{\cmsss
Z\kern-.36em Z}} {\lower1.2pt\hbox{\cmsss Z\kern-.36em
Z}}\else{\cmss Z\kern-.4em Z}\fi}


\def\ie{{\it i.e.}}
\def\eg{{\it e.g.}}

\nref\KhouryBZ{ J.~Khoury, B.~A.~Ovrut, N.~Seiberg,
P.~J.~Steinhardt and N.~Turok, ``From big crunch to big bang,''
Phys.\ Rev.\ D {\bf 65}, 086007 (2002) [hep-th/0108187].
}

\nref\SeibergHR{ N.~Seiberg, ``From big crunch to big bang - is it
possible?,'' hep-th/0201039.
}

\nref\BalasubramanianRY{ V.~Balasubramanian, S.~F.~Hassan,
E.~Keski-Vakkuri and A.~Naqvi, ``A space-time orbifold: A toy
model for a cosmological singularity,'' hep-th/0202187.
}

\nref\CornalbaFI{ L.~Cornalba and M.~S.~Costa, ``A New
Cosmological Scenario in String Theory,'' Phys.\ Rev.\ D {\bf 66},
066001 (2002) [hep-th/0203031].
}

\nref\NekrasovKF{ N.~A.~Nekrasov, ``Milne universe, tachyons, and
quantum group,'' hep-th/0203112.
}

\nref\SimonMA{
J.~Simon,
``The geometry of null rotation identifications,''
JHEP {\bf 0206}, 001 (2002)
[hep-th/0203201].
}

\nref\TolleyCV{
A.~J.~Tolley and N.~Turok,
``Quantum fields in a big crunch / big bang spacetime,''
Phys.\ Rev.\ D {\bf 66}, 106005 (2002)
[hep-th/0204091].
}

\nref\LiuFT{
H.~Liu, G.~Moore and N.~Seiberg,
``Strings in a time-dependent orbifold,''
JHEP {\bf 0206}, 045 (2002)
[hep-th/0204168].
}

\nref\ElitzurRT{
S.~Elitzur, A.~Giveon, D.~Kutasov and E.~Rabinovici,
``From big bang to big crunch and beyond,''
JHEP {\bf 0206}, 017 (2002)
[hep-th/0204189].
}

\nref\CornalbaNV{
L.~Cornalba, M.~S.~Costa and C.~Kounnas,
``A resolution of the cosmological singularity with
orientifolds,''
Nucl.\ Phys.\ B {\bf 637}, 378 (2002)
[hep-th/0204261].
}

\nref\CrapsII{
B.~Craps, D.~Kutasov and G.~Rajesh,
``String propagation in the presence of cosmological singularities,''
JHEP {\bf 0206}, 053 (2002)
[hep-th/0205101].
}

\nref\KachruKX{
S.~Kachru and L.~McAllister,
``Bouncing brane cosmologies from warped string compactifications,''
hep-th/0205209.
}

\nref\LawrenceAJ{
A.~Lawrence,
``On the instability of 3D null singularities,''
JHEP {\bf 0211}, 019 (2002)
[hep-th/0205288].
}

\nref\GordonJW{
C.~Gordon and N.~Turok,
``Cosmological perturbations through a general relativistic bounce,''
hep-th/0206138.
}

\nref\MartinecXQ{
E.~J.~Martinec and W.~McElgin,
``Exciting AdS orbifolds,''
JHEP {\bf 0210}, 050 (2002)
[hep-th/0206175].
}

\nref\LiuKB{
H.~Liu, G.~Moore and N.~Seiberg,
``Strings in time-dependent orbifolds,''
JHEP {\bf 0210}, 031 (2002)
[hep-th/0206182].
}

\nref\FabingerKR{
M.~Fabinger and J.~McGreevy,
``On smooth time-dependent orbifolds and null singularities,''
hep-th/0206196.
}

\nref\HorowitzMW{
G.~T.~Horowitz and J.~Polchinski,
``Instability of spacelike and null orbifold singularities,''
Phys.\ Rev.\ D {\bf 66}, 103512 (2002)
[hep-th/0206228].
}

\nref\BuchelKJ{
A.~Buchel, P.~Langfelder and J.~Walcher,
``On time-dependent backgrounds in supergravity and string theory,''
hep-th/0207214.
}

\nref\HemmingKD{
S.~Hemming, E.~Keski-Vakkuri and P.~Kraus,
``Strings in the extended BTZ spacetime,''
JHEP {\bf 0210}, 006 (2002)
[hep-th/0208003].
}

\nref\HashimotoNR{
A.~Hashimoto and S.~Sethi,
``Holography and string dynamics in time-dependent backgrounds,''
hep-th/0208126.
}

\nref\SimonCF{
J.~Simon,
``Null orbifolds in AdS, time dependence and holography,''
JHEP {\bf 0210}, 036 (2002)
[hep-th/0208165].
}

\nref\AlishahihaBK{
M.~Alishahiha and S.~Parvizi,
``Branes in time-dependent backgrounds and AdS/CFT correspondence,''
JHEP {\bf 0210}, 047 (2002)
[hep-th/0208187].
}

\nref\SatohNJ{
Y.~Satoh and J.~Troost,
``Massless BTZ black holes in minisuperspace,''
hep-th/0209195.
}

\nref\CaiSV{ R.~G.~Cai, J.~X.~Lu and N.~Ohta, ``NCOS and D-branes
in time-dependent backgrounds,'' Phys.\ Lett.\ B {\bf 551}, 178
(2003) [hep-th/0210206].
}

\nref\BachasQT{
C.~Bachas and C.~Hull,
``Null brane intersections,''
hep-th/0210269.
}

\nref\OkuyamaPC{
K.~Okuyama,
``D-branes on the null-brane,''
hep-th/0211218.
}

\nref\PapadopoulosBG{
G.~Papadopoulos, J.~G.~Russo and A.~A.~Tseytlin,
``Solvable model of strings in a time-dependent plane-wave background,''
hep-th/0211289.
}

\lref\GiveonTQ{ A.~Giveon and D.~Kutasov,
JHEP {\bf 0001}, 023 (2000) [hep-th/9911039].
}
\lref\WeinbergVP{ E.~J.~Weinberg and A.~Wu, ``Understanding
Complex Perturbative Effective Potentials'', Phys.\ Rev.\ D {\bf
36}, 2474 (1987).
}
\lref\GuthYA{ A.~H.~Guth and S.~Pi, ``The Quantum Mechanics Of The
Scalar Field In The New Inflationary Universe'', Phys.\ Rev.\ D
{\bf 32}, 1899 (1985).
}

\lref\MarcusVS{ N.~Marcus, ``Unitarity And Regularized Divergences
In String Amplitudes'', Phys.\ Lett.\ B {\bf 219}, 265 (1989).
}
\lref\FischlerTB{ W.~Fischler and L.~Susskind, ``Dilaton Tadpoles,
String Condensates And Scale Invariance. 2'', Phys.\ Lett.\ B {\bf
173}, 262 (1986).
}

\lref\SenMG{ A.~Sen, ``Non-BPS states and branes in string
theory,'' hep-th/9904207.
}
\lref\MaldacenaRE{ J.~Maldacena, ``The large N limit of
superconformal field theories and supergravity,'' Adv.\ Theor.\
Math.\ Phys.\  {\bf 2}, 231 (1998) [Int.\ J.\ Theor.\ Phys.\  {\bf
38}, 1113 (1998)] [hep-th/9711200].
}

\lref\WittenQJ{ E.~Witten, ``Anti-de Sitter space and
holography'', Adv.\ Theor.\ Math.\ Phys.\  {\bf 2}, 253 (1998)
[hep-th/9802150].
}

\lref\GubserBC{ S.~S.~Gubser, I.~R.~Klebanov and A.~M.~Polyakov,
``Gauge theory correlators from non-critical string theory,''
Phys.\ Lett.\ B {\bf 428}, 105 (1998) [hep-th/9802109].
}

\lref\HarveyNA{ J.~A.~Harvey, D.~Kutasov and E.~J.~Martinec, ``On
the relevance of tachyons'', hep-th/0003101.
}

\lref\AntoniadisAA{ I.~Antoniadis, C.~Bachas, J.~R.~Ellis and
D.~V.~Nanopoulos, ``Cosmological String Theories And Discrete
Inflation,'' Phys.\ Lett.\ B {\bf 211}, 393 (1988).
}

\lref\AntoniadisVI{ I.~Antoniadis, C.~Bachas, J.~R.~Ellis and
D.~V.~Nanopoulos, ``An Expanding Universe In String Theory,''
Nucl.\ Phys.\ B {\bf 328}, 117 (1989).
}

\lref\AntoniadisUU{ I.~Antoniadis, C.~Bachas, J.~R.~Ellis and
D.~V.~Nanopoulos, ``Comments On Cosmological String Solutions,''
Phys.\ Lett.\ B {\bf 257}, 278 (1991).
}

\lref\HawkingNK{
S.~W.~Hawking,
``The Chronology protection conjecture,''
Phys.\ Rev.\ D {\bf 46}, 603 (1992).
}

\lref\WittenYR{ E.~Witten, ``On string theory and black holes,''
Phys.\ Rev.\ D {\bf 44}, 314 (1991).
}

\lref\DijkgraafBA{ R.~Dijkgraaf, H.~Verlinde and E.~Verlinde,
``String propagation in a black hole geometry,'' Nucl.\ Phys.\ B
{\bf 371}, 269 (1992).
}

\lref\VenezianoPZ{ G.~Veneziano, ``String cosmology: The pre-big
bang scenario,'' hep-th/0002094.
}

\lref\BanksYP{ T.~Banks and W.~Fischler, ``M-theory observables
for cosmological space-times,'' hep-th/0102077.
}

\lref\Gradshteyn{I.~S.~Gradshteyn and I.~M.~Ryzhik, {\it Table of
Integrals, Series and Products}, Academic Press (1994).}

\lref\Abramowitz{M.~Abramowitz and I.~A.~Stegun, {\it Handbook of
Mathematical Functions}, Dover Publications, New York (1970).}

\lref\Vilenkin{ N.~J.~Vilenkin, ``Special Functions and the Theory
of Group Representations'', AMS, 1968.}

\lref\DiFrancescoUD{ P.~Di Francesco and D.~Kutasov, ``World sheet
and space-time physics in two-dimensional (Super)string theory,''
Nucl.\ Phys.\ B {\bf 375}, 119 (1992) [hep-th/9109005].
}
\lref\KutasovPV{ D.~Kutasov, ``Some properties of (non)critical
strings,'' hep-th/9110041.
}

\lref\TeschnerFT{ J.~Teschner, ``On structure constants and fusion
rules in the SL(2,C)/SU(2) WZNW model,'' Nucl.\ Phys.\ B {\bf
546}, 390 (1999) [hep-th/9712256].
}

\lref\PolchinskiRQ{ J.~Polchinski, ``String Theory. Vol. 1: An
Introduction To The Bosonic String,'' {\it  Cambridge, UK: Univ.
Pr. (1998) 402 p}. }

\lref\ColeyUH{ A.~A.~Coley, ``Dynamical systems in cosmology,''
gr-qc/9910074.
}

\lref\Abbott{ L.~F.~Abbott and S.~Deser, ``Stability Of Gravity
With A Cosmological Constant,'' Nucl.\ Phys.\ B {\bf 195}, 76
(1982).}

\lref\BirrellIX{ N.~D.~Birrell and P.~C.~Davies, ``Quantum Fields
In Curved Space,'' {\it  Cambridge, Uk: Univ. Pr. ( 1982) 340p}. }

\lref\GasperiniAK{ M.~Gasperini and G.~Veneziano, ``O(d,d)
covariant string cosmology,'' Phys.\ Lett.\ B {\bf 277}, 256
(1992) [hep-th/9112044].
}

\lref\MeissnerZJ{ K.~A.~Meissner and G.~Veneziano, ``Symmetries of
cosmological superstring vacua,'' Phys.\ Lett.\ B {\bf 267}, 33
(1991).
}

\lref\MeissnerGE{ K.~A.~Meissner and G.~Veneziano, ``Manifestly
O(d,d) invariant approach to space-time dependent string vacua,''
Mod.\ Phys.\ Lett.\ A {\bf 6}, 3397 (1991) [hep-th/9110004].
}

\lref\SmithUP{ E.~Smith and J.~Polchinski, ``Duality survives time
dependence,'' Phys.\ Lett.\ B {\bf 263}, 59 (1991).
}

\lref\SilversteinXN{ E.~Silverstein, ``(A)dS backgrounds from
asymmetric orientifolds,'' hep-th/0106209.
}

\lref\TseytlinWR{ A.~A.~Tseytlin, ``Duality and dilaton,'' Mod.\
Phys.\ Lett.\ A {\bf 6}, 1721 (1991).
}

\lref\TseytlinHT{ A.~A.~Tseytlin, ``On the form of the black hole
solution in D = 2 theory,'' Phys.\ Lett.\ B {\bf 268}, 175 (1991).
}

\lref\deAlwisAS{ S.~P.~de Alwis,
Phys.\ Lett.\ B {\bf 289}, 278 (1992) [hep-th/9205069].
}

\lref\Mathews{J.~Mathews and R.~L.~Walker, {\it Mathematical methods of
physics}, Addison-Wesley (1970).}

\lref\NappiKV{
C.~R.~Nappi and E.~Witten,
``A Closed, expanding universe in string theory,''
Phys.\ Lett.\ B {\bf 293}, 309 (1992)
[hep-th/9206078].
}

\lref\BanadosWN{
M.~Banados, C.~Teitelboim and J.~Zanelli,
``The Black Hole In Three-Dimensional Space-Time,''
Phys.\ Rev.\ Lett.\  {\bf 69}, 1849 (1992)
[hep-th/9204099].
}
\lref\BanadosGQ{
M.~Banados, M.~Henneaux, C.~Teitelboim and J.~Zanelli,
``Geometry of the (2+1) black hole,''
Phys.\ Rev.\ D {\bf 48}, 1506 (1993)
[gr-qc/9302012].
}
\lref\SenNU{
A.~Sen,
``Rolling tachyon,''
JHEP {\bf 0204}, 048 (2002)
[hep-th/0203211].
}
\lref\SenIN{
A.~Sen,
``Tachyon matter,''
JHEP {\bf 0207}, 065 (2002)
[hep-th/0203265].
}
\lref\SenAN{
A.~Sen,
``Field theory of tachyon matter,''
Mod.\ Phys.\ Lett.\ A {\bf 17}, 1797 (2002)
[hep-th/0204143].
}

\lref\CarrollAR{
S.~M.~Carroll,
``Lecture notes on general relativity,''
gr-qc/9712019.
}

\lref\FabingerKR{
M.~Fabinger and J.~McGreevy,
``On smooth time-dependent orbifolds and null singularities,''
hep-th/0206196.
}

\lref\SenVV{
A.~Sen,
``Time evolution in open string theory,''
JHEP {\bf 0210}, 003 (2002)
[hep-th/0207105].
}
\lref\SenQA{
A.~Sen,
``Time and tachyon,''
hep-th/0209122.
}
\lref\GibbonsTV{
G.~Gibbons, K.~Hashimoto and P.~Yi,
``Tachyon condensates, Carrollian contraction of Lorentz group, and  fundamental strings,''
JHEP {\bf 0209}, 061 (2002)
[hep-th/0209034].
}
\lref\FelderSV{
G.~N.~Felder, L.~Kofman and A.~Starobinsky,
``Caustics in tachyon matter and other Born-Infeld scalars,''
JHEP {\bf 0209}, 026 (2002)
[hep-th/0208019].
}

\lref\GerasimovZP{
A.~A.~Gerasimov and S.~L.~Shatashvili,
``On exact tachyon potential in open string field theory,''
JHEP {\bf 0010}, 034 (2000)
[hep-th/0009103].
}

\lref\HorowitzAP{
G.~T.~Horowitz and A.~R.~Steif,
``Singular String Solutions With Nonsingular Initial Data,''
Phys.\ Lett.\ B {\bf 258}, 91 (1991).
}
\lref\KutasovQP{
D.~Kutasov, M.~Marino and G.~W.~Moore,
``Some exact results on tachyon condensation in string field theory,''
JHEP {\bf 0010}, 045 (2000)
[hep-th/0009148].
}

\lref\KutasovAQ{
D.~Kutasov, M.~Marino and G.~W.~Moore,
``Remarks on tachyon condensation in superstring field theory,''
hep-th/0010108.
}

\lref\StromingerPC{
A.~Strominger,
``Open string creation by S-branes,''
hep-th/0209090.
}

\lref\LambertHK{
N.~D.~Lambert and I.~Sachs,
``Tachyon dynamics and the effective action approximation,''
hep-th/0208217.
}

\lref\MaldacenaKR{
J.~M.~Maldacena,
``Eternal black holes in Anti-de-Sitter,''
hep-th/0106112.
}

\lref\KounnasWC{
C.~Kounnas and D.~L\"ust,
``Cosmological string backgrounds from gauged WZW models,''
Phys.\ Lett.\ B {\bf 289}, 56 (1992)
[hep-th/9205046].
}

\lref\LustVD{
D.~L\"ust,
``Cosmological string backgrounds,''
hep-th/9303175.
}

\lref\KrausIV{ P.~Kraus, H.~Ooguri and S.~Shenker, ``Inside the
Horizon with AdS/CFT,'' hep-th/0212277.
}

\lref\HashimotoSK{ K.~Hashimoto, P.~M.~Ho and J.~E.~Wang,
``S-brane actions,'' hep-th/0211090.
}

\Title{\vbox{
\hbox{hep--th/0212215}
\hbox{WIS/46/02-DEC-DPP}
\hbox{EFI-02-45}
}} {\vbox{\centerline{Comments on Cosmological Singularities}
\vskip 10pt \centerline{in String Theory}}}
\centerline{Micha Berkooz\footnote{$^\dagger$}{Incumbent of the Recanati career
development chair for energy research}$^{,}$\footnote{$^{\ddagger}$}
{\tt Micha.Berkooz@weizmann.ac.il}$^{,a}$,
Ben Craps\footnote{$^\star$}{{\tt
craps,kutasov,rajesh@theory.uchicago.edu}}$^{,b}$,
David Kutasov$^{\star,b}$ and
Govindan Rajesh$^{\star,b}$
}
\bigskip
\centerline{\it $^a$Department of Particle Physics, The Weizmann Institute
of Science, Rehovot 76100, Israel}
\medskip
\centerline{\it $^b$Enrico Fermi Institute, University of Chicago,
5640 S. Ellis Av., Chicago, IL 60637, USA}
\bigskip
\bigskip
\noindent
We compute string scattering amplitudes in an orbifold of
Minkowski space by a boost, and show how certain divergences in
the four point function are associated with graviton exchange near
the singularity. These divergences reflect large tree-level
backreaction of the gravitational field. Near the singularity, all
excitations behave like massless fields on a 1+1 dimensional
cylinder. For excitations that are chiral near the singularity, we
show that divergences are avoided and that the backreaction is
milder. We discuss the implications of this for some cosmological
spacetimes. Finally, in order to gain some intuition about what
happens when backreaction is taken into account, we study an open
string rolling tachyon background as a toy model that shares some
features with $\Rop^{1,1}/\Zop$.


\Date{12/02}

\newsec{Introduction}
Time dependent solutions in string theory are of interest,
both for describing the early universe, and for studying
various dynamical issues. In particular, the behavior near
cosmological singularities has recently been studied by many
researchers, including~\refs{\KhouryBZ- \PapadopoulosBG}.

The first step in such studies typically involves determining
the wavefunctions of particles in the singular geometry, and
in particular the manner in which these wavefunctions are to
be continued through the singularities. This can be done \eg\
by using orbifold~\refs{\NekrasovKF,\LiuFT,\CrapsII} or coset
conformal field theory~\refs{\ElitzurRT,\CrapsII} techniques.

The second step involves an analysis of stability of the
spacetime under small perturbations. Very general arguments
lead one to believe that already classically a large backreaction
of the geometry to small perturbations is to be expected, since
any non-zero stress tensor gets infinitely amplified near a
singularity \HorowitzAP. Quantum mechanical effects generically
lead to further large backreaction \refs{\HawkingNK,\HorowitzMW}.

The classical backreaction in a class of models which can be described
near the singularity by certain Lorentzian orbifolds has recently been studied
in~\refs{\LiuFT,\LawrenceAJ,\LiuKB}. It was found that, as expected, these
orbifolds
suffer from instabilities associated with the divergent stress tensor of matter
near the singularity. This is reflected in certain new divergences of
the $2\to2$ tree level scattering amplitude of particles in the geometry.
These divergences are associated with the region near the singularity,
and signal the fact that, when backreaction is included, the curvature
and string coupling in these backgrounds grow without bound (while
without backreaction, the string coupling is weak everywhere, and the
curvature vanishes everywhere except at the singularity).

We here will examine a different class of models, which reduce
near a cosmological singularity to \eqn\aaa{\IR^{1,1}/\IZ\times
\IR^d~,} where the orbifold generator acts as a boost in
$\IR^{1,1}$. This type of singularity, which is usually referred
to in the literature as Milne or Misner spacetime, arises in a
number of examples of recent interest: \item{(1)} The Nappi-Witten
model~\refs{\NappiKV,\ElitzurRT}, in which this singularity
appears at the intersection of two copies of a closed,
big-bang/big-crunch universe, and certain non-compact static
regions (the ``whiskers'' of~\ElitzurRT). \item{(2)} The
big-bang/big-crunch cosmology described in~\CrapsII, corresponding
to a circle shrinking from finite to zero size, and then back to
the original size. \item{(3)} The spacelike singularity of a BTZ
black hole with $M>0$, $J=0$~\refs{\BanadosWN,\BanadosGQ}; see
\HemmingKD\ for a recent discussion and references. \item{(4)}
$\IR^{1,1}/\IZ$ is also of independent interest, and has been
studied as such in \refs{\NekrasovKF,\TolleyCV} (in the latter
paper with the Rindler wedges omitted).

The plan of the paper is as follows. In section~2 we review
the structure of the orbifold \aaa, the wavefunctions on it,
and the relation between the wavefunctions and the choice of
vacuum for quantum fields on $\IR^{1,1}/\IZ$.

In section~3 we perform a string calculation of the tree level
$2\to 2$ scattering amplitude in the spacetime \aaa, and
study its singularities. We are particularly interested in
singularities of the scattering amplitude that are associated
with the cosmological singularity (the fixed set of the boost
action).

In section~4 we compare the results of the string calculation to
a gravity analysis of the same amplitude. This helps one identify
the part of the amplitude associated with the backreaction to the
growing stress tensor of perturbations near the orbifold singularity.
We find that in general the dilaton remains finite
near the singularity, while the curvature of the metric (including
backreaction) blows up and causes the divergence found in section~3.
We also find that one can fine-tune the initial conditions in such a way
that the large backreaction is avoided.

The physical picture is the following. All quantum fields on \aaa\ behave
near the singularity like massless fields living on a $1+1$ dimensional
cylinder. Left-moving excitations carry a large amount of $T_{++}$,
while right-moving ones have a large $T_{--}$. Large backreaction
occurs when both left and right movers are present. It is associated
with processes in which left and right movers on the cylinder collide
near the singularity. The fine-tuning referred to in the previous paragraph
corresponds in this language to a situation where only left-movers or
only right-movers are present near the singularity. One then has
a wavefront moving with the speed of light to the left (or right)
and no violent collisions/backreaction take place. General arguments
suggest that in this case the solution is well behaved---\eg\
$\alpha'$ and $g_s$ corrections to the original background are small.

In section~5 we discuss some implications of the analysis of
sections 3 and 4 to some of the systems mentioned earlier. In
particular, we discuss the backreaction for different choices of
vacuum in Milne spacetime. While in the vacuum inherited from the
underlying Minkowski spacetime, the backreaction is always large,
in another natural choice of vacuum one often finds a small
backreaction. In the Nappi-Witten model, the backreaction to the
modes studied in~\ElitzurRT\ is small. In the model of~\CrapsII\
one finds a large backreaction. We also briefly discuss the case
of the non-rotating BTZ black hole; for a more comprehensive
recent discussion see \KrausIV.

In situations where the classical backreaction is large, it would be
interesting to understand what happens to the background when backreaction
is taken into account.
To gain insight into this interesting problem, we discuss in section~6
an open string toy model, which exhibits some of the features described
for $\IR^{1,1}/\IZ$ above. The toy model is the dynamics of open strings
on an unstable D-brane, in the background of a homogenous rolling
tachyon \SenNU.

We show that an effective field theory, which should provide a good description
of the dynamics of the tachyon at late times \SenAN, exhibits features similar
to those found for the closed string systems in sections~3 -- 5. Generic
perturbations lead to ``large backreaction'' on the tachyon background
at late times. By fine-tuning the solution in a way similar to
that in section~4, one can arrange for the backreaction to be small.
Moreover, by thinking of the D-brane as a collection of $D0$-branes,
the fine-tuning in question is precisely that needed to keep the $D0$-branes
at rest relative to each other at late times. The large backreaction in the
general case is associated with $D0$-branes approaching each other and
interacting in a non-trivial way via open strings stretched between
them. We discuss the role of these interactions in the full open string
problem, and the possible implications for the cosmological singularities
studied in sections~3 -- 5.

In section~7 we summarize the results, and discuss some
open issues. Three appendices contain some useful technical results.

\newsec{Wavefunctions and vertex operators in $\Rop^{1,1}/\Zop$}
\subsec{Geometry of $\Rop^{1,1}/\Zop$} In $D$-dimensional
Minkowski space, \eqn\metricMink{
ds^2=-(dX^0)^2+(dX^1)^2+\cdots+(dX^{D-1})^2~, } define
\eqn\xpm{\eqalign{X^\pm&=(X^0\pm X^1)/\sqrt2~;\cr \vec
X&=(X^2,\ldots,X^{D-1})~, } } so that the metric reads \eqn\Mink{
ds^2=-2dX^+dX^-+d\vec X^2~. } $\Rop^{1,1}/\Zop$ is obtained by
orbifolding with the group $\Zop$ generated by the boost
\eqn\boost{ X^\pm\mapsto e^{\pm2\pi}X^\pm~. } The resulting
spacetime consists\foot{We neglect one dimensional pieces of
spacetime that come from moding out the $X^+X^-=0$ locus.} of four
cones (times $\Rop^{D-2}$, which we will suppress in some of the
formulae below), touching at the spacelike singularity $X^\pm=0$.
We will refer to the four cones as the ``early time region'' or
``past Milne wedge'' ($X^+,X^-<0$), the ``late time region'' or
``future Milne wedge'' ($X^+,X^->0$), and the ``regions with
closed timelike curves'' or ``Rindler wedges'' ($X^+X^-<0$). In
the early and late time regions, it is useful to define
coordinates $(t,x)$, \eqn\tx{ X^\pm={1\over\sqrt2}te^{\pm x}~, }
in terms of which the metric and identification are
\eqn\metrid{\eqalign{ ds^2&=-dt^2+t^2dx^2~;\cr x&\sim x+2\pi~. } }
It will also be convenient to define the conformal time coordinate
$\eta$ by (\eg\ in the early time region) \eqn\conftime{
t=-e^\eta~, } so that the metric becomes \eqn\metrconf{
ds^2=e^{2\eta}(-d\eta^2+dx^2)~. }

\subsec{Wavefunctions} Wavefunctions on the orbifold
\aaa\ are wavefunctions on Minkowski space
that are invariant under \boost. A convenient basis of
wavefunctions is given by~\NekrasovKF \eqn\Nekr{ \psi(X^+,X^-,\vec
X)_{p^+,p^-;l}={e^{i{\vec p}\cdot{\vec X}}\over 2\sqrt2\pi i}
\int_\Rop dw\,e^{i(p^+X^-e^{-w} +p^-X^+e^{w}+lw)}~,\ \ \
l\in\Zop~. } The normalization has been chosen such that the
Klein-Gordon norm is 1. For $p^+,p^-\geq0$ (which we will assume
in what follows), these wavefunctions are superpositions of
negative frequency plane waves in Minkowski space, so they
describe excitations over the (adiabatic) vacuum inherited from
Minkowski space (see \eg\ \BirrellIX). The mass shell condition is
$2p^+p^-=m^2$ ($m$ being the two-dimensional mass, which includes
contributions from momenta in any additional dimensions), so by
shifting $w$ and multiplying the wavefunction by a phase we can
put \eqn\ppm{p^+=p^-=m/\sqrt2~.} Using the integral
representation~\Gradshteyn \eqn\hankel{ H^{(1)}_\nu(\tilde
z)={1\over i\pi}e^{-\half i\nu\pi}\int_0^\infty dy\,e^{\half
i\tilde z(y+{1\over y})}\,y^{-\nu-1} } of the Hankel function
$H^{(1)}_\nu$, valid for $0<\arg {\tilde z}<\pi$ or for $\arg
{\tilde z}=0, -1<{\rm Re}\;\nu<1$, the wavefunction \Nekr\ can be
brought to the form (for $X^+,\, X^-\geq0$)
\eqn\NekrHankel{\eqalign{ \psi_{m,l}=&{1\over2\sqrt2\pi
i}\Bigl({p^+X^-\over p^-X^+}\Bigr)^{il\over2} \int_0^\infty
dy\,e^{i\sqrt{p^+p^-X^+X^-} (y+{1\over y})}\, y^{il-1} \cr
=&{1\over2\sqrt2}e^{l\pi\over2} \Bigl({p^+X^-\over
p^-X^+}\Bigr)^{{il\over2}} H^{(1)}_{-il}(\tilde z)\cr =&
{1\over2\sqrt2}e^{l\pi\over2}e^{-ilx}H^{(1)}_{-il}(mt)~, } } where
we have used \eqn\ztildenu{ \tilde
z\equiv2\sqrt{p^+p^-X^+X^-}=mt~. } We see that $l$ is the momentum
along the $x$ circle \metrid. The expression of the wavefunction
\Nekr\ in the other regions of $\Rop^{1,1}/\Zop$ is described in
appendix~C.

The Hankel function $H^{(1)}_\nu(\tilde z)$ can be written as
\eqn\hankelbessel{ H^{(1)}_{-il}(\tilde z)=
-{1\over\sinh(l\pi)}(e^{-l\pi}J_{-il}(\tilde z)-J_{il}(\tilde z))~,
}
where
the Bessel function $J_{-il}(\tilde z)$ has the power series expansion
\eqn\powerbessel{\eqalign{
J_{-il}(\tilde z)&=\Bigl({\tilde z\over2}\Bigr)^{-il}\sum_{k=0}^\infty
{\Bigl(-{\tilde z^2\over4}\Bigr)^k\over (k!)\Gamma(-il+k+1)}\cr
&={1\over\Gamma(-il+1)}\Bigl({\tilde z\over2}\Bigr)^{-il}
\sum_{k=0}^\infty
{\Bigl(-{\tilde z^2\over4}\Bigr)^k\over
(k!)(1-il)(2-il)\cdots(k-il)}~.
}}
Note that
\eqn\lminusl{
e^{l\pi\over2}H^{(1)}_{-il}(\tilde z)=e^{-l\pi\over2}H^{(1)}_{il}(\tilde z)~,
}
so that $\psi_{-l}(t,x)=\psi_{l}(t,-x)$; in particular, both are superpositions
of negative frequency waves in Minkowski space. Their complex conjugates, which
are superpositions of positive frequency modes in Minkowski space, can be
written in terms of $H^{(2)}_{il}=(H^{(1)}_{-il})^*$. These positive frequency
modes annihilate the vacuum inherited from Minkowski space.

Another basis of solutions to the wave equation is given by the functions
(again in the region where $X^+,\, X^-\geq 0$)
\eqn\besselwf{
\chi_{m,l}(t,x)={1\over2\sqrt{\sinh(\pi |l|)}}J_{i|l|}(mt)e^{-ilx}
}
together with their complex conjugates. For $t\to 0$ ($\eta\to -\infty$),
 \besselwf\ become purely negative frequency
with respect to conformal time $\eta$, as can be seen using
\powerbessel: \eqn\chiexp{ \chi_{m,l}(t,x)\sim
{1\over2\sqrt{\sinh(\pi |l|)}}{1\over\Gamma(1+i|l|)}
\Bigl({m\over2}\Bigr)^{i|l|}e^{i|l|\eta}e^{-ilx}~. } As a
consequence, the complex conjugate modes annihilate a state which
in the limit $m=0$ is called the conformal vacuum \BirrellIX. This
is to be contrasted with the modes \NekrHankel, which for $t\to 0$
involve both positive and negative frequencies with respect to
$\eta$: \eqn\Nekrleading{ \psi_l\sim {1\over 2\sqrt{2}\sinh(\pi
l)}\Bigl[-\Bigl({me^{\eta + x}\over {2}} \Bigr)^{-il}{e^{-{\pi
l\over2}}\over\Gamma(1-il)}+\Bigl({me^{\eta-x}\over {2}}
\Bigr)^{il}{e^{{\pi l\over2}}\over\Gamma(1+il)}\Bigr]~. } Using
the identity \eqn\normgamma{
|\Gamma(1+il)|^2=\Gamma(1+il)\Gamma(1-il)={\pi l\over\sinh(\pi
l)}~ } and writing $\Gamma(1+il) = e^{i\phi_l}\sqrt{\pi
l\over\sinh(\pi l)}$, with $\phi_l = -\phi_{-l}$, we can
rewrite~\Nekrleading\ as \eqn\Nekrleadtwo{ \psi_l\sim {1\over
2\sqrt{2\pi l \sinh(\pi l)}} \Bigl[-\Bigl({me^{\eta + x}\over{2}}
\Bigr)^{-il}e^{-{\pi l\over2} +
i\phi_l}+\Bigl({me^{\eta-x}\over{2}} \Bigr)^{il}e^{{\pi l\over2} -
i \phi_l}\Bigr]~. } We see that, as mentioned in the introduction,
wavefunctions of all particles behave near the Milne singularity
like those of massless $1+1$ dimensional fields on the cylinder
labelled by $(\eta, x)$. We will return to this fact later.

Finally, one might want to consider wavefunctions which are
Fourier transforms of \Nekr: \eqn\newwave{ \Psi(\sigma) = \sum_l
\psi_l e^{-il\sigma}~.} Since $l$ is the momentum along the circle
labelled by $x$ \metrid, $\sigma$ can be thought of as a position
along this circle. Using~\Nekr, we have \eqn\newwavetwo{
\Psi(\sigma) = \sum_n e^{{im\over\sqrt{2}}(X^- e^{-(\sigma + 2\pi
n)} + X^+ e^{\sigma + 2\pi n})}~, } so $\Psi(\sigma)$ is a plane
wave in Minkowski space superposed with all its images under the
orbifold group. As $\eta\to-\infty$, $\psi_l$, and hence $\Psi$,
splits into left and right-moving parts, $\Psi = \Psi^- + \Psi^+$
where \eqn\Psipm{ \Psi^{\pm}(\sigma) = \sum_n a^{\pm}_n(\sigma)
e^{in(x\pm \eta)}~.} The coefficients $a^{\pm}_n(\sigma)$ can be
extracted from \Nekrleadtwo: \eqn\apmdef{ a^{\pm}_n(\sigma)  =
{\mp 1\over 2\sqrt{2\pi n\sinh(\pi n)}} \Bigl({m\over
{2}}\Bigr)^{\pm in} e^{\pm{ \pi n\over 2} \mp i\phi_n + i
n\sigma}~.} The Klein-Gordon norm of these wavefunctions diverges,
so we will only consider \Nekr\ and \besselwf\ in this paper.

\newsec{$2\to 2$ scattering in string theory}

We are interested in the behavior of strings near the singularity
$X^\pm=0$ of the orbifold \boost. Therefore we will compute
classical string scattering amplitudes involving the vertex
operators~\Nekr, following~\LiuFT\ who studied a different time-dependent
orbifold. We will focus on the $2\to 2$ scattering amplitude,
since this is the simplest case in which gravitational backreaction
is expected to play a role.

Two and three point functions are studied in appendix~A.
It turns out that although the three point functions exhibit an interesting
structure, this structure is associated to the asymptotic regions
in the Milne wedges, not to the singularity. A very similar
behavior was found in~\LiuFT.

In the present section, we compute the four point function of the tachyon
vertex operators \Nekr. We will study the process $1+2\to3+4$. The
corresponding amplitude is
\eqn\fourptrep{\eqalign{
&\langle\psi^*_3\psi^*_4\psi_1\psi_2\rangle= {1\over 64\pi^4}
\int_{-\infty}^\infty
dw_1\cdots dw_4e^{i\sum \epsilon_i l_i w_i}
\langle\prod_{i=1}^4e^{i({\epsilon_i m_i\over\sqrt{2}}X^-e^{- w_i}+
{\epsilon_i m_i\over\sqrt{2}}X^+e^{w_i}+ \epsilon_i \vec p_i\cdot\vec X)}\rangle\cr
&={(2\pi)^{20}\over 4}\int dX^+dX^-\int dw_1\cdots dw_4 e^{i\sum \epsilon_i l_i w_i}
e^{i{X^-\over\sqrt{2}}\sum \epsilon_i m_i e^{-w_i}
+i{X^+\over\sqrt{2}}\sum \epsilon_i m_i e^{w_i}}\times\cr
&\hskip30pt\times\delta^{24}(\sum \epsilon_i\vec p_i)
\int d^2z|z|^{p_1\cdot p_3}|1-z|^{p_1\cdot p_4}~. \cr} }
The coefficients $\epsilon_i$ are $1$ for the incoming particles 1 and 2, and
$-1$ for the outgoing particles 3 and 4. The mass shell condition is
(setting $\alpha'=1$)
\eqn\massshell{m^2 = - 4 + \vec p^2~.}
As before, $m^2$ is the effective two-dimensional mass squared,
and we assume it to be positive, so we can define $m$ to be positive
as well.
Define
\eqn\vi{
v_i = e^{w_i - w_1},~~i = 2,3,4~.
}
Then the Mandelstam invariants are given by
\eqn\stu{\eqalign{
s&=-(p_1+p_2)^2=-8+m_1m_2(v_2+{1\over v_2}) -2\vec p_1\cdot \vec p_2~;\cr
t&=-(p_1-p_3)^2=-8-m_1m_3(v_3+{1\over v_3}) +2\vec p_1\cdot \vec p_3~;\cr
u&=-(p_1-p_4)^2=-8-m_1m_4(v_4+{1\over v_4}) +2\vec p_1\cdot \vec p_4~.
}
}
We can reduce the expression for the four-point function to a single integral
as follows. We first perform the $z$ integral\foot{The $z$ integral runs,
as usual, over the (Euclidean) worldsheet -- the sphere or plane. This
might seem like a problem since we are studying an inherently Minkowski
signature spacetime, \aaa, and thus should take the worldsheet to have
Minkowski signature as well.
However, the Euclidean calculation is only used here to arrive at the
Shapiro-Virasoro amplitude, which can be taken to be the starting
point of the analysis, and used directly in Minkowski spacetime.}
in \fourptrep:
\eqn\zint{\int d^2z|z|^{p_1\cdot p_3}|1-z|^{p_1\cdot p_4} = 2\pi
{\Gamma(-1 - {s\over 4})\Gamma(-1 - {t\over 4}) \Gamma(-1 -
{u\over 4})\over \Gamma(2 + {s\over 4}) \Gamma(2 + {t\over
4})\Gamma(2 + {u\over 4}) }~.}
Defining $G(x) = {\Gamma(-1 - {x\over 4})\over \Gamma(2 + {x\over 4})}$
and performing the $w_1$ and $X^\pm$
integrals, we can write the four point function as
\eqn\fourptc{\eqalign{ {(2\pi)^{24}\over 4}\delta^{24}
\left(\sum \epsilon_i \vec p_i\right)
\delta\left(\sum \epsilon_i l_i\right)
&\int^{\infty}_{0} dv_2 dv_3 dv_4\; G(s) G(t)
G(u)\times\cr &\times \delta\left(m_1 + \sum_{j=2}^4 \epsilon_j m_j
v_j\right) \delta\left(m_1 + \sum_{j=2}^4 {\epsilon_j m_j \over v_j}\right)
\prod_{j=2}^4 v_j^{i \epsilon_j l_j - 1}~.} }
We now perform the $v_2,v_3$ integrals. Setting the arguments of
the delta functions to zero amounts to solving
\eqn\vroots{\eqalign{ &m_1 + m_2 v_2 - m_3 v_3 - m_4 v_4  = 0~;\cr
&m_1 + {m_2 \over v_2} - {m_3 \over v_3} - {m_4\over v_4}  = 0\cr }}
for $v_2,v_3$.
The solutions are
\eqn\sols{\eqalign{ v_2&={AB+m_2^2-m_3^2\mp\sqrt\Delta\over 2m_2B}~;\cr
v_3&=- {AB+m_3^2-m_2^2\pm\sqrt\Delta\over 2m_3B}~, }}
where
\eqn\delt{\eqalign{
A&=-m_1 + m_4 v_4~;\cr
B&=-m_1 + {m_4\over v_4}~;\cr
\Delta&=(m_2^2-m_3^2)^2-2AB(m_2^2+m_3^2)+A^2B^2~. }}
Note that $v_2,v_3$ should be positive, so one should
retain only the positive solutions among \sols.
Including a Jacobian factor
\eqn\Jac{{1\over|m_2m_3|}{v_2^2v_3^2\over|v_2^2-v_3^2|} } from the delta
functions and plugging in the solutions \sols,
the four point function can be reduced to the following single integral:
\eqn\fourptd{
\sum {(2\pi)^{24}\over 4}\delta^{24}\left(\sum \epsilon_i \vec p_i\right)
\delta\left(\sum \epsilon_i l_i\right) \int_0^\infty dv_4\;
G(s) G(t) G(u) {{v_2}^{i l_2 + 1} {v_3}^{-il_3 + 1}
v_4^{- i l_4 - 1}\over |m_2 m_3 ({ v_2}^2 - { v_3}^2)|}~, }
where the sum runs over the positive
solutions among \sols.
We are interested in the divergences of the four-point function, since they can
potentially teach us something about the singularity.

Consider the $v_4\to\infty$ region of the integral. In this limit the
positive solution among \sols\ (corresponding to the upper sign)
is given by
\eqn\vapprox{{v}_2\approx {m_4 v_4\over m_2}~,\ \ {v}_3\approx {m_3\over m_1}~,
}
so that the Mandelstam variables $t$ and $s$ are
\eqn\tsol{
t \approx - (\vec p_1 - \vec p_3)^2~, \ \ s \approx m_1 m_4 v_4~.
}
This is the Regge limit ${s\to\infty}$, $t$ fixed, in which
\eqn\virsap{G(s) G(t) G(u)\to
-\Bigl({s\over4}\Bigr)^{2 + {t\over 2}} {\Gamma[-1 - {t\over 4}]\over \Gamma[2 + {t\over 4}]}~.
}
The $v_4\to\infty$ limit of the four-point function is therefore
\eqn\fourpte{\eqalign{ - {(2\pi)^{24}\over 2^{6-(\vec p_1 - \vec p_3)^2}}&
\delta^{24}
\left(\sum \epsilon_i\vec p_i\right) \delta\left(\sum \epsilon_i l_i\right)
(m_1m_4)^{1-\half(\vec p_1 - \vec p_3)^2}
\left({m_4\over m_2}\right)^{i l_2}
\left({m_3\over m_1}\right)^{- i l_3}\times\cr
&\times{\Gamma[-1 + {(\vec p_1 - \vec p_3)^2\over 4}]\over
\Gamma[2 - {(\vec p_1 - \vec p_3)^2\over 4}]}
\int dv_4 \; v_4^{-\half(\vec p_1 - \vec p_3)^2 + i (l_2 - l_4)}~.}}
The integral over $v_4$ diverges from $v_4\to\infty$
whenever $(\vec p_1 - \vec p_3)^2 \le 2$.
The $v_4\to 0$ limit is equivalent to the $v_4\to\infty$ limit
(they are complex conjugates).

In the above expression, we have set $\alpha'=1$.
For later comparison with the gravity calculation, it is also
useful to consider the limit $\alpha'\to 0$, or more precisely
the limit $\alpha' t\to0$ ($\alpha's$ and $\alpha'u$ cannot go
to zero at the same time since $s+t+u=-16/\alpha'$; this is an
irrelevant complication, which is due to the fact that the mass
of the tachyon is of order the string scale. Similar results would
be obtained for fields with masses well below the string scale).
Near $x=0$, we have
$\Gamma[x-1]\sim - {1\over x}$, so that the $\alpha'\to 0$ limit
of the four-point function in the Regge limit becomes
\eqn\fourptf{{(2\pi)^{24}\over16}\delta^{24}\left(\sum \epsilon_i \vec p_i\right)
\delta\left(\sum \epsilon_i l_i\right)
\left({m_1m_4\over(\vec p_1 - \vec p_3)^2}\right)
\left({m_4\over m_2}\right)^{i l_2}
\left({m_3\over m_1}\right)^{- i l_3}
\int dv_4 \; v_4^{i (l_2 - l_4)}~.}

There are additional divergences from other regions of the integral \fourptd.
Some of them may be understood as different versions of the above. For example,
$m_2^2 < m_3^2$, $m_1^2 < m_4^2$ and
$B$ a small positive number corresponds to
large $s$, fixed $u$. In this regime the four-point function
diverges whenever $(\vec p_1 - \vec p_4)^2 \le 2$.

A different kind of divergence occurs when $v_2=v_3$ and
$(m_2-m_3)^2 = (m_1-m_4)^2$.
This is an IR effect which is not associated with the
singularity (see appendix~B).

\newsec{Gravity analysis}

In this section we will show that the divergence
\fourptf\ is due to exchange of gravitons near the
singularity, and signals a large gravitational
backreaction in that region. We will also see that
in some situations the backreaction is milder than for
generic kinematics, and the tree level $2\to 2$ scattering
amplitude is finite.

In subsection {\it 4.1} we will compute the massless exchange
contribution to the four point function \fourptrep\ in an alternative way,
and show that, in a certain kinematic regime, the dominant contribution
to \fourptf\ comes from graviton exchange near the singularity. In
subsection {\it 4.2} we will study the backreaction in dilaton-gravity
coupled to a scalar field, and relate these calculations to
the scattering amplitudes studied earlier.

\subsec{An alternative calculation of the $2\to 2$ scattering amplitude
in the $\alpha' t\to 0$ limit}

In the $\alpha't\to0$ limit, with $t$ a Mandelstam variable \stu,
\zint\ can be replaced by
\eqn\tzeropole{
 2\pi
{\Gamma(-1 - {s\over 4})\Gamma(-1 - {t\over 4}) \Gamma(-1 -
{u\over 4})\over \Gamma(2 + {s\over 4}) \Gamma(2 + {t\over
4})\Gamma(2 + {u\over 4}) }\rightarrow-{2\pi(p_1\cdot p_2)^2\over t}~.
}
In this limit, \zint, and as a result the four point function
studied in section~3, is dominated by a massless exchange in the
$t$-channel.
Thus, computing the $2\to2$ scattering of the wavefunctions \Nekr\
(with $t$ small in string units) amounts to computing
\fourptrep, with the last integral
replaced by the r.h.s. of \tzeropole:
\eqn\fourptgrav{\eqalign{
&\langle\psi^*_3\psi^*_4\psi_1\psi_2\rangle=
{(2\pi)^{20}\over 4}\int dX^+dX^-\int dw_1\cdots dw_4 e^{i\sum \epsilon_i l_i w_i}
e^{i{X^-\over\sqrt{2}}\sum \epsilon_i m_i e^{-w_i}
+i{X^+\over\sqrt{2}}\sum \epsilon_i m_i e^{w_i}}\cr
&\hskip30pt\times\delta^{24}(\sum \epsilon_i\vec p_i)
\Bigl(-{2\pi(p_1\cdot p_2)^2\over t}\Bigr)~.
}
}
One way of computing \fourptgrav\ is to repeat the analysis of section~3,
\ie\ first perform the integrals over $X^\pm$, which give delta functions of
momentum conservation on the covering space, and then integrate over $w_i$.
The result is the $\alpha't\to0$ limit of \fourptd, which reduces in the
$v_4\to \infty$ limit to \fourptf. Note that in this limit one has
$t\simeq -(\vec p_1-\vec p_3)^2$ (see \tsol).
Also, to relate \fourptgrav\ to \fourptf\ one uses $s \approx m_1 m_2
v_2$ (see \vapprox\ and \tsol); this amounts to
$p_1\cdot p_2\approx-p_1^+p_2^-$. These observations will be relevant for
comparison with a second way of computing \fourptgrav, to which we turn
next.

Another way of computing \fourptgrav\ is to perform the $w_i$ integrals
first. Then one obtains
\eqn\fourptwx{
\langle\psi^*_3\psi^*_4\psi_1\psi_2\rangle=
-(2\pi)^{24}\delta^{24}(\sum \epsilon_i\vec p_i)2\pi\int dX^+dX^-
\partial_\mu\psi_{m_1,l_1}\partial_\nu\psi^*_{m_3,l_3} {1\over\partial^2}
\partial^\mu\psi_{m_2,l_2}\partial^\nu\psi_{m_4,l_4}^*.
}
The large $v_4$ divergences found in section 3 come from the term
\eqn\fourptwxbis{\eqalign{
\langle\psi^*_3\psi^*_4\psi_1\psi_2\rangle&\simeq \cr
&-(2\pi)^{24}\delta^{24}(\sum \epsilon_i\vec p_i)2\pi\int dX^+dX^-
\partial_-\psi_{m_1,l_1}\partial_-\psi^*_{m_3,l_3} {1\over\partial^2}
\partial_+\psi_{m_2,l_2}\partial_+\psi_{m_4,l_4}^*
}
}
(see the previous paragraph). Now consider
the contribution to \fourptwxbis\ from the wedge $X^\pm>0$; the
other three wedges are discussed in detail in appendix~C. We
will see shortly that for a certain range of the
transverse momenta, the
integral \fourptwxbis\ is dominated by the region of small
$X^+X^-$. Let us assume for now
that the integral \fourptwxbis\ is dominated by the
region of small $X^+X^-$, and that it is consistent to use the leading
behavior  \Nekrleading\ of the wavefunctions,
\eqn\NekrleadingX{
\psi_{m,l}\sim {1\over 2\sqrt{2}\sinh(\pi l)}\Bigl[-\Bigl({mX^+\over\sqrt2}
\Bigr)^{-il}{e^{-{\pi l\over2}}\over\Gamma(1-il)}+\Bigl({mX^-\over\sqrt2}
\Bigr)^{il}{e^{{\pi l\over2}}\over\Gamma(1+il)}\Bigr]~.
}
Under these assumptions, the $1/\partial^2$ operator in \fourptwxbis\
acts only on the $\exp(i\epsilon_j\vec p_j\cdot \vec X)$ parts of the
full wavefunctions. Thus, one has
\eqn\oneoverbox{
{1\over\partial^2}=-{1\over(\vec p_2-\vec p_4)^2}
}
and \fourptwxbis\ becomes
\eqn\fourptwxtris{\eqalign{&
{(2\pi)^{24}\delta^{24}(\sum \epsilon_i\vec p_i) 2\pi \Bigl({m_2\over\sqrt2}
\Bigr)^{-il_2}
\Bigl({m_4\over\sqrt2}\Bigr)^{il_4}\Bigl({m_1\over\sqrt2}\Bigr)^{il_1}
\Bigl({m_3\over\sqrt2}\Bigr)^{-il_3}
e^{{\pi\over2}(l_1+l_3-l_2-l_4)}l_1l_2l_3l_4
\over 64 (\prod\sinh(\pi l_i)) \Gamma(1-il_2)\Gamma(1+il_4)
\Gamma(1+il_1)\Gamma(1-il_3)(\vec p_1-\vec p_3)^2}\cr&
\times\int_0^\infty{dX^+dX^-\over(X^+X^-)^2}(X^+)^{i(l_4-l_2)}(X^-)^{i(l_1-l_3)}~.
}
}
Using $X^{\pm} = {1\over\sqrt{2}}e^{\eta\pm x}$, we can write the integral
in the above expression as
\eqn\Cnewintaa{\eqalign{
\int_0^{\infty} dX^+\int_0^{\infty} dX^-
&{(X^+)}^{i(l_4-l_2)-2}{(X^-)}^{i(l_1-l_3)-2} =\cr
&2 \int_{-\infty}^{\infty} d\eta\int_{-\infty}^{\infty} dx
\left(e^{\eta}\over\sqrt{2}\right)^{i(l_1+l_4-l_2-l_3)-2}
e^{ix(l_3+l_4-l_1-l_2)}~.}}
Performing the $x$ integral, and defining $v=2e^{-2\eta}$,
we can simplify this to
\eqn\Cnewintab{
2\pi\delta(l_1+l_2-l_3-l_4)\int_0^{\infty}dv\, v^{i(l_2+l_3-l_1-l_4)/2}~.}
In appendix~C, we show that upon adding the contributions of
the other three wedges of spacetime, \fourptwxtris\ reproduces
\fourptf, with $v$ playing the role of $v_4$. Thus, contributions from
large $v_4$ (or large Mandelstam variable $s$ with fixed $t$) correspond to
contributions from the region near the singularity.

There is a small subtlety in the preceding discussion, which we
would like to mention at this point. In \oneoverbox, we assumed
that in evaluating the $1/\partial^2$ operator, we can use the
asymptotic form of the wavefunctions $\psi_{m,l}$ near the
singularity $X^+ X^-=0$. In fact, if one is very close to the
t-channel graviton pole, the exchanged particle can propagate for
a large distance and the amplitude is {\it not} dominated by the
behavior of the wavefunctions near the singularity. However, one
can show that if the momentum transfer $(\vec p_2-\vec p_4)^2$ is
small compared to the string scale, but large compared to the
two-dimensional masses squared, the approximation leading to
\oneoverbox\ is valid at large $v_4$, and that subleading
corrections in $m^2/(\vec p_2-\vec p_4)^2$ correspond to
subleading corrections in $1/v_4$ in the analysis of section 3.

Thus, we conclude that the divergence \fourptf\ of the four point
function is associated with exchange of massless particles near
the singularity.
An interesting by-product of this analysis is that it
points to particular kinematical situations where this divergence
is absent. Consider two incoming particles 1 and 2 whose
wavefunctions are purely left moving (or purely right moving) near
the singularity, \ie\ they only depend on $X^+$ (or only on
$X^-$).
Then it is clear that the contribution \fourptwxbis\ responsible
for the divergence vanishes. Examples
of wavefunctions that are chiral near
the singularity are given by \besselwf. In section~5, we discuss
some situations in which they are physically relevant.

\subsec{Backreaction in dilaton gravity}

In this subsection we would like to compute the classical
backreaction of the graviton and dilaton to an incoming
tachyon perturbation, and relate it to the amplitude calculations
of the previous subsection.

Consider the action
\eqn\acgrav{S=\int d^D x\sqrt{-g}e^{-2\Phi}\left(
R+4g^{\mu\nu}\partial_\mu\Phi\partial_\nu\Phi-
g^{\mu\nu}\partial_\mu T\partial_\nu T-m^2T^2\right)~,}
where
$g_{\mu\nu}$ is the string frame metric, $\Phi$ the dilaton,
and $T$ a scalar field of mass $m$. It is convenient to transform
to Einstein frame, by defining
\eqn\einstmet{\tilde g_{\mu\nu}=e^{-{4\Phi\over D-2}}g_{\mu\nu}~.}
The action is now
\eqn\aceins{S=\int d^D x\sqrt{-\tilde g}\left(
\tilde R-{4\over D-2}\tilde g^{\mu\nu}\partial_\mu\Phi\partial_\nu\Phi-
\tilde g^{\mu\nu}\partial_\mu T\partial_\nu T-m^2T^2e^{4\Phi\over D-2}\right)~.}
In this action, $\Phi$ is essentially decoupled from $\tilde g$ and so we
can treat them separately. The equation of motion for $\Phi$, to leading order
in $\Phi$, is
\eqn\eomphi{\partial^2\Phi=\half m^2 T^2~.}
Thus, we see that the dilaton is only sensitive to the mass term, and
in particular, it does not seem to diverge for the wavefunctions we consider
in this paper (\eg,
if the two $T$'s on the r.h.s. of \eomphi\ are combinations of
$(X^\pm)^{\mp il_j}$, see \NekrleadingX ).

Now we turn to a
discussion of the backreaction of the metric. We expand
\eqn\smallfield{\tilde g_{\mu\nu}=\eta_{\mu\nu}+h_{\mu\nu}~.}
Plugging into the action \aceins\ we get
\eqn\ssmall{S=\int d^Dx\left\{-\half\left(
\partial_\mu h^{\mu\nu}\partial_\nu h-\partial_\rho h^{\rho\sigma}\partial_\mu
h^\mu_\sigma+\half\partial_\mu h^{\rho\sigma}\partial^\mu h_{\rho\sigma}-
\half\partial_\mu h\partial^\mu h\right)+h^{\mu\nu} T_{\mu\nu}\right\}~,}
where Lorentz indices are raised and lowered with the flat metric $\eta_{\mu\nu}$
and we defined $h\equiv {h^\mu}_\mu$. The stress tensor that enters the
action is
\eqn\stressten{T_{\mu\nu}=\partial_\mu T\partial_\nu T-\half
\eta_{\mu\nu}\left[(\partial T)^2+m^2 T^2\right]~.}

We would like to fix the de Donder gauge
\eqn\gaugefix{\partial_\mu h^{\mu\nu}=\half\partial^\nu h~.}
We do this in a way analogous to the way Feynman gauge is introduced
in QED. We add to the Lagrangian a term
\eqn\gaugfeyn{-\half(\partial_\mu h^{\mu\nu}-\half\partial^\nu h)^2~,}
which leads to the following form of the action:
\eqn\sdedonder{S=\int d^Dx \left\{-{1\over4}
\partial_\mu h^{\rho\sigma}\partial^\mu h_{\rho\sigma}+
{1\over8}\partial_\mu h\partial^\mu h+h^{\mu\nu} T_{\mu\nu}\right\}~.}
Varying with respect to $h_{\mu\nu}$, we find the equation of motion
\eqn\hmunu{\partial^2 h_{\mu\nu}=-2\partial_\mu T\partial_\nu T
-{2\over D-2} m^2 T^2\eta_{\mu\nu}~.}
It is useful in verifying this to write also the
trace of this equation:
\eqn\hhh{\partial^2 h=-2(\partial T)^2-{2D\over D-2} m^2 T^2~.}
Note that this gives the Ricci tensor of the perturbed metric,
since in the gauge \gaugefix\
\eqn\ccuurr{R_{\mu\nu}=-\half\partial^2 h_{\mu\nu}=\partial_\mu T
\partial_\nu T+{1\over D-2}\eta_{\mu\nu} m^2 T^2}
(see \eg\ \refs{\CarrollAR}, eq.~(6.6)).
In particular, we see that there is large backreaction
when the field $T$ has both $X^+$ and $X^-$ dependent pieces
near the singularity. If the field $T$ is chiral, say only
a function of $X^+$, then only $R_{++}$ will be large near the
singularity, and the problem seems much milder behaved (for instance,
powers of the Ricci scalar are finite, and $\alpha'$ corrections
to the Einstein action constructed out of powers of the Ricci tensor
can be neglected). This is in
agreement with the observation made at the end of subsection~{\it 4.1},
that divergences of four point functions are absent if the
incoming fields are chiral.

To study the effect of the backreaction \eomphi, \ccuurr\ on the scalar
field $T$, one classically integrates out $h_{\mu\nu}$, $\Phi$, by
plugging the solutions of their equations of motion back into the classical
action \aceins. To order $T^4$, one finds
\eqn\efffourpt{S_4=\int d^Dx\left\{\partial_\mu T\partial_\nu T{1\over \partial^2}
\partial_\mu T\partial_\nu T
-\half\left[(\partial T)^2+m^2 T^2\right]{1\over \partial^2}
\left[(\partial T)^2+m^2 T^2\right]\right\}~.}
The second term on the r.h.s. of \efffourpt\ is despite
appearances non-singular.
By integrating by parts, one can show that for on-shell tachyons it equals
$-T^2\partial^2 T^2/8$. The first term is of the form \fourptwx, which
was used to compute the $2\to 2$ scattering amplitude of $T$'s in
subsection {\it 4.1}.

\newsec{Applications}

The picture emerging from the analysis in sections~3 and~4 is that
generic small perturbations of the Milne orbifold \aaa\
lead to large classical gravitational
backreaction, and that this is reflected in divergences in four
point functions of these perturbations.
The backreaction is milder if the perturbations
are chiral near the singularity, \ie, if the incoming wavefunctions
of particles 1 and 2 depend only on
$X^+$ or only on $X^-$ close to the singularity. The
divergences in the four point function associated with the
singularity are absent for such fine-tuned perturbations, as we
mentioned at the end of subsection~{\it 4.1}. In this section we will
discuss some qualitative implications of this observation for a few
spacetimes that look locally like the Milne orbifold \aaa.

\subsec{Milne orbifold}

When quantizing fields on the Milne orbifold, one has to choose
a vacuum. One natural choice of vacuum is the one
inherited from the Minkowski space prior to orbifolding.
Excitations of this vacuum are described by the
wavefunctions \NekrHankel, which near the singularity involve
both chiralities (see \Nekrleading). Thus, the analysis of
sections~3 and~4 implies that this vacuum
exhibits large classical backreaction to any perturbation.

On the other hand, as we discussed around \chiexp, in another
natural vacuum state (see \eg\ \refs{\BirrellIX}) the positive
energy
wavefunctions are given by \besselwf. Near the singularity, they
depend only on $X^-$ for positive momentum $l$, and only on $X^+$
for negative momentum. So we conclude that in this state, there
is large backreaction if one has incoming particles of both
positive and negative momentum. Large backreaction is
avoided if all incoming particles are moving in the same
direction on the cylinder near the singularity (and similarly
for outgoing particles).

\subsec{Big crunch/big bang cosmology of \CrapsII}

In \CrapsII, an orbifold of a coset CFT was studied which
describes a spacetime with a big crunch/big bang singularity, and
with a number of asymptotic regions as well as compact regions with closed
timelike curves (see also \refs{\KounnasWC,\LustVD}).
Close to the big crunch/big bang singularity, the
spacetime looks like a Milne orbifold \aaa.
Natural in and out vacua were identified, and
the amount of particle creation was computed in string theory.
It was found that, due to the presence of different asymptotic regions and
in particular the singularities connecting them,
particle creation of any given mode
did not decay with energy for
large energies, unlike the situation in smooth spacetimes
where it is know to decay
exponentially with energy. This raises the suspicion that there
should be large backreaction in this model.

The modes annihilating the natural incoming vacuum of \CrapsII\ turn out
to involve both chiralities near the Milne
singularity, so there is large backreaction to any
perturbation of this vacuum. It would be interesting to see if and
how this is related to the large amount of particle creation in
this model found in \CrapsII.

\subsec{Nappi-Witten model}

In \ElitzurRT, a coset CFT was studied which describes a four-dimensional
spacetime
containing a few copies of a closed big-bang/big-crunch universe,
which was originally studied in \NappiKV,
as well as a number of non-compact static regions
which extend to spatial infinity (``whiskers'').
The closed cosmological regions are attached to the whiskers
at a singularity, which looks locally like the Milne orbifold
\aaa.

As discussed in \ElitzurRT, it is natural to study scattering amplitudes
of $n$ to $m$ particles in a given whisker. The incoming state corresponds to
particles sent in from infinity in the whisker. Their wavefunctions are given
by (3.29) in \ElitzurRT; they have the property that near the
Milne singularity (see fig.~3 in \ElitzurRT) they are chiral in region
1 (the whisker) and they vanish in the cosmological region I.
Similarly, the wavefunctions of the outgoing states are chiral
with the opposite chirality near the Milne singularity in region
1, and vanish in cosmological region II.

Thus, the analysis of sections~3 and~4 leads to the conclusion that
the contribution from the whisker to $n\to m$ scattering amplitudes is
finite (classically),
while the contribution to these amplitudes from the closed cosmological
regions vanishes. In fact, one can obtain these amplitudes by analytically
continuing the correlation functions from Euclidean space.

Actually, amplitudes in whisker $1$ are not unitary in this case.
The situation is analogous to that in eternal black hole spacetimes.
The full geometry contains a second whisker (denoted by~$1'$ in
fig.~3 of \ElitzurRT), connected to the original one at the Milne
singularity. Thus, there is another asymptotic region where
information can go -- spatial infinity in whisker~$1'$. The
situation is similar to that described in \MaldacenaKR\ for
black holes (see also \HemmingKD). The full geometry contains two disconnected
boundaries, at infinity in regions~$1$ and~$1'$, on which asymptotic
states are defined. Amplitudes in whisker~$1$ can be computed using a density
matrix, corresponding to tracing out the degrees of freedom in whisker~$1'$.
One can also compute correlations
between whiskers~$1$ and~$1'$, which are non-zero because the
states of the two
whiskers are entangled.

The amplitudes of the modes described above do not get contributions from the
cosmological big bang/big crunch regions.
Thus, it is not very surprising that they do not give rise to large
backreaction of the geometry. Generic incoming modes correspond to
particles coming in from infinity in whisker~$1$, as well as from the
cosmological region~$I$. Such modes correspond to wavefunctions
which contain both chiralities near the Milne singularity, and thus
lead to large backreaction. More generally, amplitudes that probe
dynamics in the compact, cosmological regions of spacetime are
expected to suffer from the divergences discussed in sections~3 and~4,
from one or both of its big bang and big crunch Milne singularities.

\subsec{BTZ black hole}

As mentioned in the introduction, non-rotating BTZ black holes
have a spacelike singularity of Milne type, and one might wonder
whether our analysis sheds any light on its fate in string theory.

In asymptotically AdS spacetimes, one is interested in computing
boundary correlations functions, which are the AdS analogues of S-matrix
elements. Thus, the question is whether these boundary correlation
functions are sensitive to physics near the BTZ singularity.
A natural way of defining boundary correlation functions is by analytic
continuation from Euclidean space. Euclidean BTZ spacetime does not
contain a singularity, and the boundary correlation functions on it
are well behaved. One natural continuation from Lorentzian to
Euclidean BTZ involves continuing ``Schwarzschild time''
$t_{\rm sch}\to i\theta$. This maps the Euclidean black hole manifold
to a single region outside the horizon of the black hole
(say, region I in fig. 1 of \MaldacenaKR).
The Euclidean correlation
functions are mapped under this continuation to correlation
functions\foot{These correlation functions are not unitary
since the CFT on the boundary of region I is entangled
with one living
on the boundary of region II, and the latter has been traced over
in computing them.} of insertions on the boundary of region I.
This continuation is clearly
insensitive to physics near the singularity, since only the behavior
of wavefunctions in region I enters. Thus, our analysis is irrelevant
for it.

Another continuation from Minkowski to Euclidean BTZ involves a
continuation of Kruskal time $t_{\rm kruskal}\to i\theta$. In this
case, the Euclidean black hole is mapped to the full extended
Lorentzian BTZ spacetime. Euclidean boundary correlation functions
map to Lorentzian correlation functions with insertions on both
boundaries (in regions I and II). Assuming that these boundary
correlation functions probe local physics in the full extended BTZ
spacetime, and in particular near the singularity, they can be
used to resolve the singularity; see \MaldacenaKR\ and \KrausIV\
for recent discussions.

From the perspective of our discussion here, this second
continuation is more puzzling. The wavefunctions (in the
Hartle-Hawking state) that one gets in this continuation (\eg\ eq.
(2.7) in \MaldacenaKR) diverge near the BTZ singularity like
\eqn\ffff{\phi\sim\log(X^+X^-)} in the coordinates used in
sections 2 -- 4. Thus, the contribution to these amplitudes from
the vicinity of the singularity is divergent, as in the discussion
of sections~3 and~4 above. This divergence signals a large
backreaction of the metric, as in section~4. Since the full
amplitude obtained by continuation from Euclidean spacetime is
finite, it must be that from the point of view of the discussion
in sections~3 and~4, the divergences due to different
singularities cancel \KrausIV. However, this is a non-local
cancellation, and there is some tension between the statement that
it occurs in all correlation functions, and the expectation that
boundary correlation functions can be used to probe local physics
in the bulk of the full extended BTZ spacetime. Also, it is not
completely clear in what sense one can neglect the backreaction
near a particular singularity, which appears to be large. This
issue clearly requires a better understanding.

\newsec{Open string toy model}

In this section we will discuss an open string model which shares some features
with the closed string systems discussed in the previous sections. Consider a
D-string in bosonic string theory.\foot{A similar analysis can be performed for
non-BPS branes and brane-antibrane systems in type II string theory.}
As is well known, the lowest lying excitation of the D-string is tachyonic.
The condensation of this open string tachyon, $T$, leads to the
disappearance of the brane. Condensation of spatially
dependent modes of the tachyon leads to lower dimensional D-branes,
which are also unstable and can further decay by tachyon condensation.

An interesting background of the theory on the D-brane
is one in which the tachyon (which is taken to be
constant along the D-string) starts at early time $x^0\to-\infty$
at the top of its potential, $T=0$, which corresponds to the
original D-brane, and evolves at late time ($x^0\to\infty$) to the
bottom of its potential, $T=\infty$.
The late time solution is not just the ``no brane'' (closed string vacuum)
solution, since the energy of the original brane does not disperse in the
classical approximation, but rather is stored in the kinetic energy
of the open string field. This leads to a pressureless fluid ``tachyon matter''
state \refs{\SenNU,\SenIN}.

In conformal field theory on the strip (\ie\ classical open string
theory), this ``rolling tachyon'' background is described by adding
the boundary interaction
\eqn\boundint{\delta S=\int d\tau e^{x^0(\tau)}}
to the worldsheet Lagrangian of a D-brane\foot{More precisely,
\boundint\ is the form of the interaction at early times, $x^0
\to-\infty$, where the perturbation is small. The precise form
of the perturbation at large $x^0$ depends on the renormalization
prescription.}. This boundary CFT has not been analyzed in
detail (see \StromingerPC\ for a recent discussion), but it has
been argued that the endpoint of the time evolution is highly
unstable; for example, if one turns on an arbitrarily small
coupling to closed strings, one expects the tachyon matter
to decay into closed strings and disappear.

A closer analogue of the instabilities of cosmological spacetimes
to small perturbations at early times discussed in previous section,
would in this case be instabilities of the rolling tachyon background
\boundint\ to small
open string perturbations at early times. We will next see that
such instabilities indeed do arise and discuss their physical
interpretation.

We will use a field theoretic effective action which has been argued
\SenAN\  to give a good description of tachyon dynamics for large $T$.
It is useful for our purposes, since this is the region which
we are most interested in.

The action is
\eqn\actionsen{\eqalign{
S&=\int dxdx^0\,{\cal L}~;\cr
{\cal L}&=-V(T)\,\sqrt{1+\eta^{\mu\nu}\partial_\mu T\partial_\nu T}~,
}}
where the potential is taken to be
\eqn\vt{
V(T)=e^{-T}~,
}
and the signature of the metric is $\eta={\rm diag}(-,+)$.
The potential \vt\ does not have a maximum unlike the actual potential
of the tachyon field in open string theory. It can be thought of as
describing the evolution of the tachyon for energies much smaller than
the energy of the original D-brane. The conclusions of our analysis
would presumably be similar for other potentials which go exponentially
to zero at large $T$, such as those that arise in boundary string field theory
\refs{\GerasimovZP,\KutasovQP,\KutasovAQ} (see \refs{\LambertHK, \GibbonsTV,
\SenQA} for discussions of more general potentials).

The energy density corresponding to \actionsen\ is
\eqn\enden{
T_{00}={e^{-T}(1+ T'^2)\over\sqrt{1-\dot T^2+T'^2}}~.
}
In terms of the momentum conjugate to $T$,
\eqn\conj{
\Pi(x)={\delta S\over\delta(\dot T(x))}={e^{-T}\dot T\over
\sqrt{1-\dot T^2+T'^2}}~,
}
the Hamiltonian is
\eqn\hamil{\eqalign{
H&=\int dx\,T_{00}~;\cr
T_{00}&=\sqrt{(\Pi^2+e^{-2T})(1+T'^2)}
}
}
and the Hamilton equations of motion are
\eqn\eom{\eqalign{
\dot\Pi&=\partial_x\bigl({T'\sqrt{\Pi^2+e^{-2T}}\over
\sqrt{1+T'^2}}\bigr)+{e^{-2T}\sqrt{1+T'^2}\over
\sqrt{\Pi^2+e^{-2T}}}~;\cr
\dot T&={\Pi\sqrt{1+T'^2}\over\sqrt{\Pi^2+e^{-2T}}}~.
}
}
A homogeneous solution of the equations of motion is given by
\eqn\homsol{\eqalign{
T_0&={\rm log}[{1\over E}\cosh (x^0)]~;\cr
\Pi_0&=E\tanh (x^0)~,
}
}
which for late times approaches
\eqn\homsollate{\eqalign{
T_0&\sim x^0~;\cr
\Pi_0&\sim E~.
}
}
Note, in particular, that the leading asymptotic behavior of $T$ is independent
of the energy density $E$, while that of $\Pi$ does depend on $E$. More generally, there
is a class of solutions of the equations of motion that goes for late times like  \SenAN
\eqn\nonhomsollate{\eqalign{
T&\sim x^0~;\cr
\Pi&\sim f(x)~,
}
}
where $f(x)$ is an arbitrary function of the spatial coordinates, and the corrections
to \nonhomsollate\ are exponentially small at late times. Eq.\ \hamil\ implies that
the energy density of such solutions is $T_{00}=|f(x)|$. In fact, it has been argued
in \FelderSV\ that the most general solution of the equations of motion \eom\ approaches
at late times a solution of the first order equation
\eqn\firsor{\dot T^2-T'^2=1~.}
The solutions \nonhomsollate\ are indeed of this form.

We would like next to perform a classical stability analysis of the homogenous
rolling tachyon solution \homsol. Thus, we expand the tachyon field $T$ as
$T= T_0 + \phi$, where $T_0$ is the homogenous solution \homsol\ and $\phi$
a small fluctuation (for early times). We would like to check whether $\phi$ remains
a small perturbations as $t\to\infty$, or whether it grows to dominate the
solution.

Expanding the action \actionsen\ to second order in $\phi$, we
find \eqn\flucaction{\eqalign{ {\cal L} = E \Biggl[- {1\over
\cosh^2(x^0)} &+ {1\over 2}\left(\cosh^2(x^0) \dot\phi^2 -
\phi'^2\right) + \left({\phi\over\cosh^2(x^0)} +
\tanh(x^0)\dot\phi\right)\cr &- \left({\phi^2\over
2\,\cosh^2(x^0)} + \tanh(x^0)\phi\dot\phi\right)\Biggr]~.\cr} }
The order $\phi^0$ term is \eqn\zeropt{ \int dx^0{
-E\over\cosh^2(x^0)}  = -2E} (times the length of space, which we
suppress -- the problem is translation invariant in $x$). The
linear term in $\phi$, as well as the last term in brackets in
\flucaction\ are total derivatives and can be neglected. The
quadratic effective action for $\phi$ is thus \HashimotoSK
\eqn\phieffact{ S = {E\over 2}\int dxdx^0 \left(\cosh^2(x^0)
\dot\phi^2 - (\vec\nabla\phi)^2\right)~,} which leads to the
equations of motion \eqn\phieom{ {d\over
dx^0}\left(\cosh^2(x^0)\dot\phi\right) - \phi'' = 0~.}
Substituting $u=\tanh(x^0)$, and $\phi(u,\vec{x})= e^{ipx}F(u)$
the equation of motion simplifies to \eqn\phieoma{ (1-u^2) F''(u)
+ p^2 F(u) = 0~.} The solutions are hypergeometric functions:
\eqn\Fhgf{\eqalign{ F_1(u) &= F\left({- 1 - \sqrt{1+4p^2}\over 4},
{- 1 + \sqrt{1+4p^2}\over 4}, {1\over 2}, u^2\right)~;\cr F_2(u)
&= u\, F\left({1 - \sqrt{1+4p^2}\over 4}, {1 + \sqrt{1+4p^2}\over
4}, {3\over 2}, u^2\right)~.\cr }} We are particularly interested
in the large time behavior of the solution (\ie\ in the behavior
as $u\to 1$). The two solutions approach constants,
\eqn\Fhgflim{\eqalign{ F_1\to C_1& =
{\Gamma(\half)\over\Gamma\left({3+\sqrt{1+4p^2}\over 4}\right)
\Gamma\left({3-\sqrt{1+4p^2}\over 4}\right)}~;\cr F_2\to C_2 &=
{\Gamma({3\over 2})\over \Gamma\left({5+\sqrt{1+4p^2}\over
4}\right) \Gamma\left({5-\sqrt{1+4p^2}\over 4}\right)}~.}} $C_1$
and $C_2$ are generically non-zero, which implies that the
approximation that $\phi$ is a small perturbation (in general)
fails. Indeed, if $F$ approaches a constant at late times, the
second term in \phieffact\ makes an infinite contribution to the
action, which overwhelms the $0$'th order term \zeropt, and one
can check that the expansion of the square root that led from
\actionsen\ to \flucaction\ is not justified in this case (at late
times). Thus, generically, small perturbations create ``large
backreaction'' to the original solution \homsol, much like in the
discussion of sections~3 and~4. Another similar aspect of the two
problems is that like there we can choose a linear combination of
the two solutions \Fhgf, for which the ``backreaction'' is small.
Indeed, the linear combination \eqn\Fcomb{F(u) = C_2 F_1(u) - C_1
F_2(u)} goes to zero exponentially at late times, and for it the
quadratic action \phieffact\ is finite and the perturbative
expansion \flucaction\ is well behaved. Thus, for roughly half of
the possible initial states, we find mild backreaction.

All this is consistent with the general picture presented in \FelderSV\ and
in equation
\firsor\ above. For the fine-tuned initial data sets corresponding to \Fcomb, the
late time behavior is of the form \nonhomsollate, with $f(x)$ determined by the
initial conditions. This late time behavior corresponds to a small
perturbation of the original solution \homsollate. Generic initial conditions
lead to a generic solution of \firsor. In particular, it is no longer
true that the tachyon behaves like $T\sim x^0$ at late times, and the
perturbation is not small there. Moreover, the authors of \FelderSV, who analyzed the
general solutions of \firsor, showed that for generic initial conditions,
the solutions
develop caustics and become ill defined beyond a certain finite (but late)
time, which depends on the initial conditions. This is another aspect that is
reminiscent of the problem of backreaction near a cosmological singularity.

It is natural to ask what is the origin of the large backreaction for
generic initial data, and what is the fate of the system once it is
taken into account. We will next propose a possible physical
picture which explains these phenomena, leaving a more complete
understanding for future work.

Specifying generic initial data at early times corresponds to perturbing
the solution \homsol\ in a non-homogeneous way. In order to think about
the time evolution of the perturbations, it is convenient to split the evolution
into two steps.

At a first step, the $D1$-brane decomposes into a set of
$D0$-branes. For example, the tachyon profile\foot{This profile was recently
studied in \SenVV.}
\eqn\intach{T(x, x^0)=e^{\omega(p)x^0}\cos px}
describes \HarveyNA\ a process in which the $D$-string decomposes into
equidistant $D0$-branes at rest relative to each other. Each $D0$-brane
has a tachyon living on it, and in the solution \intach\ these tachyons
all grow uniformly. This is due to the high symmetry of the
background \intach; more generally, the $D0$-branes are located at
arbitrary points $x_i$, have general relative velocities, and the tachyons
on them are evolving in different ways.

The second step of the time evolution corresponds to the dynamics of
the $D0$-branes. At this stage of the evolution, one expects the situation
to depend on the relative motion of the $D0$-branes. If the $D0$-branes are
at rest relative to each other, the late time dynamics should be well behaved.
The $D0$-branes evolve independently from each other, with the tachyon on
each growing as time goes by, but no collisions between different $D0$-branes
taking place. This corresponds to solutions of the equations of motion of the
effective action \actionsen\ with the late time behavior \nonhomsollate.
The function $f(x)$ describes the density of $D0$-branes.

More generally, the $D0$-branes are in relative motion, and the late time
dynamics is more complicated. In particular, when two $D0$-branes
approach each other, the strings that connect them become light and
can no longer be integrated out. We believe that this is the origin of
the singularities associated with caustics found in \FelderSV, and the
growth of fluctuations found earlier in this section.
The effective action \actionsen\ appears to be well suited for describing
the collective dynamics of well separated $D0$-branes; it breaks down
when the interactions between different $D0$-branes become important
at late times.

It should also be pointed out in this context that the analysis of caustics
in \FelderSV\ used the method of characteristics to study solutions of
\firsor. In this method, the solution is obtained by following the behavior
of certain worldlines of massive particles; caustics arise when different
worldlines collide. The resulting picture is very reminiscent of what
one would get by analyzing the motion of the $D0$-branes out of which
the $D1$-brane is composed. It would be interesting to make the connection
more precise.

Armed with a qualitative understanding of the origin of the singularities associated
with generic perturbations of the solution \homsol, one can ask how these
singularities are resolved in the full open string theory. As we saw,
to do that one has to take into account the interaction between nearby
$D0$-branes. In particular, the fact that the $D0$-branes can form
bound states is not taken into account in the description \actionsen.
The analysis of \FelderSV\ seems to suggest that for generic initial
conditions, inhomogeneities in the tachyon field grow with time and
give rise to localized clusters of $D0$-branes. These bounds states of
$D0$-branes are quantum mechanical objects that need to be analyzed in the
full $D0$-brane quantum mechanics. We will leave a more detailed
analysis of this for future work.

Finally, one can ask what the open string example teaches us about
closed string dynamics near cosmological singularities. Qualitatively,
we see that the two problems are very similar. The effective action
\actionsen\ seems to play a role similar to that played by the dilaton
gravity action \acgrav\ in the gravity analysis of section 4. In the closed
string problem, the backreaction is mild when the incoming particles are
all moving with the same velocity on the cylinder near the singularity.
In the open string problem, the backreaction is mild when all the $D0$-branes
which make up the D-string are moving with the same velocity at late times.

When $D0$-branes approach each other, one can no longer ignore
effects due to open strings stretched between them. The analogous objects in the
cosmological Milne singularity are twisted strings, which become
light near the singularity and can no longer be neglected (see \BachasQT\
for related comments). These twisted strings give rise to interactions
between the incoming particles, the outcome of which has not been
analyzed so far. A natural guess is that these interactions are a first
step in the process of creation of black holes out of these colliding
particles, as in \HorowitzMW. This is similar to the creation of
bound states of $D0$-branes in the time evolution of the tachyon
in the open string problem. In fact, the analysis of the tachyon
Lagrangian \actionsen\ in \FelderSV\ gives rise to a picture quite
reminiscent of the time evolution of matter under the influence of gravity.
The nonlinearity of the equations of motion has a similar effect
in both cases: initial inhomogeneities are magnified, and the matter
tends to cluster in different places in space.

\newsec{Summary and discussion}

We have computed string scattering amplitudes in the presence of a
spacelike orbifold singularity \aaa, and found divergences similar to
those of \LiuFT. We argued that these divergences are due to
graviton exchange near the singularity, and that they reflect
large tree level gravitational backreaction. Interestingly,
divergences can be avoided for special perturbations which behave
like chiral two-dimensional fields near the singularity, and we
discussed the extent to which such special perturbations are
natural in some cosmological and black hole models. We also
briefly discussed an open string rolling tachyon model, which
seems to share some features with the cosmological backgrounds
studied earlier, and might help understand the backreaction near
the singularity.

It would be interesting to refine and extend this work in various
directions, including: \item{(1)} We have argued that for fields
that are chiral near the singularity, the gravitational
backreaction is milder. Indeed, only the $(++)$ component of the
Ricci tensor blows up, and four point functions are free of the
usual divergences associated to the singularity. The same is
expected to be true for higher point functions. It would be
interesting to find an exact CFT description of the orbifold with
chiral perturbations, and study its properties. \item{(2)} The
system discussed in section~6 would be worth understanding in more
detail. In particular, we proposed to think of the time evolution
of inhomogeneous tachyon profiles in terms of the dynamics of a
collection of D0-branes, and pointed out that this is very
reminiscent of the worldlines of auxiliary massive particles used
in \FelderSV\ to solve the equations of motion of an effective
field theory. It would be interesting to try and make this analogy
more precise, and understand the late time behavior of the system.
\item{(3)} It would of course be interesting to understand the
fate of the Milne singularity in cases where large backreaction
occurs. We have made some qualitative observations at the end of
section~6, and it would be nice if they could be made more
precise. \item{(4)} It would be interesting to obtain a better
understanding of the backreaction near the singularity of a BTZ
black hole. \item{(5)} The analysis of this paper was entirely
classical. It would be interesting to extend it to the quantum
level, at least in cases where the classical backreaction is
small.

\bigskip
\noindent{\bf Acknowledgements:}
We would like to thank P.~Kraus, F.~Larsen, H.~Liu, E.~Martinec,
W.~McElgin, G.~Moore, S.~Sethi and R.~Wald
for discussions and correspondence.
B.C.~would like to thank the Aspen Center for Physics and the Michigan Center for
Theoretical Physics for hospitality while this work was in
progress. D.K.~thanks the
Weizmann Institute and the Rutgers high energy theory group for hospitality.
This work is supported in part by the Israel-U.S.\ Binational
Science Foundation, the IRF Centers of Excellence program, the
European RTN network HPRN-CT-2000-00122, the Minerva foundation,
and by DOE grant DE-FG02-90ER40560 and NSF grant PHY-9901194.
\appendix{A}{Two and three-point functions in $\Rop^{1,1}/\Zop$.}
In the text, we focused on a particular divergence of the
four point function \fourptrep\ of the operators \Nekr, because,
as we have argued, this divergence is directly associated with the
singularity of the Milne orbifold. However, there are other
divergences in three and four point functions of \Nekr, which can be
interpreted as infrared divergences. For completeness, we discuss
two and three point functions in the present appendix, and
infrared divergences in four point functions in appendix B.
\subsec{Two point function}
The two point function of two wavefunctions \Nekr\ is computed as follows:
\eqn\twopoint{\langle \psi^*_{1} \psi_{2}\rangle= {1\over 2} \int
dw_1\,dw_2\,e^{i(-l_1w_1+l_2w_2)}\,\delta(-p_1^+e^{-w_1}+
p_2^+e^{-w_2})\delta(-p_1^-e^{w_1}+p_2^-e^{w_2})~.}
We assume $p_1^+,p_1^-,p_2^+,p_2^->0$. Using \ppm, and writing $w_1=w_++w,\
w_2=w_+-w$, this becomes
\eqn\twopointbis{ 2\int dw_+\,e^{i(-l_1+l_2)w_+}
\int dw\,e^{-i(l_1+l_2)w}\, \delta(-m_1e^{-w}+m_2e^{w})
\delta(-m_1e^{w}+m_2e^{-w})~. }
The first integral gives $2\pi\delta_{l_1,l_2}$ if we only integrate over
values of $w_+$ that cannot be identified by the action of \boost\ on the
wavefunctions. (Otherwise it would give $2\pi\delta(l_1-l_2)$.)
We finally obtain
\eqn\twopointtris{ \langle \psi^*_{1} \psi_{2}\rangle=
 {2\pi\over(m_1+m_2)}\delta_{l_1,l_2}\delta(m_1-m_2) ~.}
We now would like to know which region of spacetime gives the
dominant contribution to \twopointtris.
Rewrite \twopoint\ as
\eqn\twomoda{\eqalign{ &\langle \psi^*_{1} \psi_{2}\rangle= {1\over 8\pi^2}
\times\cr
&\int dX^+ dX^- \int dw_1\,dw_2\,e^{i(-l_1w_1+l_2w_2)}\,
e^{iX^-(-p_1^+e^{-w_1}+p_2^+e^{-w_2})}
e^{iX^+(-p_1^-e^{w_1}+p_2^-e^{w_2})}~. }}
We first perform the $w_1$, $w_2$ integrals:
\eqn\twomod{ \langle \psi_1^* \psi_2\rangle=
{1\over 8} e^{(l_1+l_2){\pi\over2}}
\int dX^+ dX^- \left({p_1^+X^-\over p_1^-X^+}\right)^{-\half i l_1}
\left({p_2^+X^-\over p_2^-X^+}\right)^{\half i l_2}
\bigl(H_{-il_1}^{(1)}(\tilde z_1)\bigr)^*H_{-il_2}^{(1)}(\tilde z_2)~, }
where, as in \ztildenu,
$\tilde z_i=2\sqrt{p^+_ip^-_iX^+X^-}$. (The expression \twomod\ is
accurate in the region $X^\pm>0$.) We now claim that the
result \twopointtris\ comes from the asymptotic regions
$X^+X^-\to\infty$ of the past and future Milne wedges. In order to
argue for this, we first concentrate on the future Milne wedge $X^\pm>0$
and replace the Hankel functions in \twomod\ by their asymptotic
expressions for large values of the argument:
${H_{\nu}(z) \sim {\sqrt{2\over \pi z} e^{i(z- \pi\nu/2 -
\pi/4)}}}$. The justification for our claim will be that this procedure
reproduces the correct result \twopointtris. Using \tx, \ppm\ and
\ztildenu, the contribution of the future wedge to \twomod\ becomes
\eqn\twomilne{ {1\over4\pi \sqrt{m_1m_2}}
\int{dx\; e^{-{ix(l_2-l_1)}}}\;\; \int_0^\infty{dt\;
e^{-i(m_1-m_2) t} }~.}
Adding the analogous contribution from the past Milne wedge
amounts to extending the range of the $t$ integral to the whole
real line. The result is
\eqn\twomil{  {\pi\over \sqrt{m_1m_2}} \delta_{l_1,l_2}
\delta(m_1-m_2) =  {2\pi\over(m_1+m_2)}
\delta_{l_1,l_2} \delta(m_1-m_2)~,}
in agreement with \twopointtris.
We conclude that the dominant contribution to the two-point function
comes from the $t\to\pm\infty$ asymptotics of the wavefunctions in the Milne
wedges. This is in agreement with the fact that the wavefunctions
\Nekr\ decay exponentially in the Rindler wedges. The result of
this appendix should be contrasted with that of section~4 and
appendix~C, where a divergence is analyzed that comes from the
singularity. In that computation, the dominant contribution comes
from the region of spacetime near the singularity, and the four
wedges make a comparable contribution.
\subsec{Three point function}
The three point function reads
\eqn\threepointa{\eqalign{
\langle \psi^*_{1} \psi_{2}\psi_{3}\rangle&=
{1\over 2 \sqrt{2} \pi i}
\int dw_1\,dw_2\,dw_3\,
\Bigl(e^{i(-l_1w_1+l_2w_2+l_3w_3)}\times\cr
&\times\delta(-m_1e^{-w_1}+m_2e^{-w_2}
+m_3e^{- w_3}) \delta(-m_1e^{w_1}+m_2e^{
w_2} +m_3e^{w_3})\Bigr)~. }}
where we have used \ppm. The $w_1$ integral gives
\eqn\threepointc{\eqalign{{ 1\over 2 \sqrt{2} \pi i}&\int dw_2\,dw_3\,
\Bigl(e^{i(l_2w_2+l_3w_3)}m_1^{-il_1}(m_2e^{- w_2}
+m_3e^{- w_3})^{-1+il_1}\times\cr
&\times\delta(-{m_1^2\over m_2e^{- w_2} +m_3e^{-
w_3}}+m_2e^{ w_2} +m_3e^{ w_3}) \Bigr)~. }}
Writing $w_2=w_++w,\ w_3=w_+-w$, this becomes
\eqn\threepointd{\eqalign{
-i\sqrt{2}\; \delta_{l_1,l_2+l_3}&\int
dw\, \Bigl(e^{i(l_2-l_3)w}m_1^{-il_1}(m_2e^{- w}
+m_3e^{ w})^{-1+il_1}\times\cr
&\times\delta({-m_1^2+m_2^2+m_3^2+m_2m_3(e^{2 w}+
e^{-2 w})\over m_2e^{- w}+m_3e^{ w}})
\Bigr)~. }}
The $w$ integral gives rise to a sum of
\eqn\threepointe{
-i\sqrt{2}\; \delta_{l_1,l_2+l_3}\,
{e^{i(l_2-l_3)w}m_1^{-il_1}(m_2e^{- w}
+m_3e^{ w})^{il_1}\over |8m_2m_3\,
{\rm sinh}(w) {\rm cosh}(w)|} }
over the roots $w=\pm w_0$ ($w_0\geq0$) of the argument of the delta function:
\eqn\roots{ \sinh^2(w_0)={m_1^2-(m_2+m_3)^2\over
4m_2m_3}~. }
We finally obtain the following expression for
the three point function with $m_1>m_2+m_3$ (the three point
function vanishes if $m_1<m_2+m_3$):
\eqn\threepoint{{- i\,\delta_{l_1,l_2+l_3}\,(e^{i\phi}+
e^{i\phi'}) \over\sqrt{2} \sqrt{m_1^2-(m_2+m_3)^2}\,
\sqrt{m_1^2-(m_2-m_3)^2}}~. }
Here $e^{i\phi}$ and $e^{i\phi'}$ are phases:
\eqn\phases{\eqalign{ e^{i\phi}&=e^{i(l_2-l_3)w_0}\,\Bigl({m_2e^{-
w_0}+m_3e^{w_0} \over m_1}\Bigr)^{il_1}~; \cr
e^{i\phi'}&=e^{-i(l_2-l_3)w_0}\,\Bigl({m_2e^{w_0}
+m_3e^{-w_0}\over m_1}\Bigr)^{il_1}~. }}
For later convenience, let us note that when $m_1 \approx m_2 +
m_3$, the three-point function can be simplified, because in this
limit, $w_0 \approx 0$:
\eqn\thptlim{ \langle \psi^*_{1}
\psi_{2}\psi_{3}\rangle\approx
{-i\,\delta_{l_1,l_2+l_3} \over
2 \sqrt{m_1m_2m_3}\sqrt{m_1-m_2-m_3}}~. }

To see where this non-analyticity in the masses comes from, it is
useful to analyze the three point
function using the explicit form of the wavefunctions in terms of
Hankel functions. We have
\eqn\threemod{\eqalign{ \langle
\psi^*_{1} \psi_{2}\psi_{3}\rangle&= {1\over 16\sqrt{2}}
e^{(l_1+l_2+l_3){\pi\over2}}
\times\cr &\times\;\;\int dX^+ dX^-
\left({X^-\over X^+}\right)^{{i\over2}(l_2+l_3-l_1)}
H_{-il_1}^{(1)}(\tilde z_1)^*H_{-il_2}^{(1)}(\tilde z_2)
H_{-il_3}^{(1)}(\tilde z_3)\cr} }
where we have used \ppm. (This expression is accurate in the
future Milne wedge.)

Let us study the contribution from the asymptotic regions in the
Milne wedges, as we did for the two point function in the
previous subsection.
Replacing the Hankel function by its asymptotics in the future Milne
wedge, the contribution of this wedge to \threemod\ becomes
\eqn\thmoda{ {e^{-\pi i/4}\over
8 \pi^{3\over 2}\sqrt{m_1m_2m_3}}\;\; \int dx e^{-
i(l_2 + l_3 - l_1) x} \;\; \int {dt\over\sqrt{t}}\;
e^{-i(m_1-m_2-m_3)t}~.}
The $t$ integral can be performed by rotating the integration
contour, clockwise or counterclockwise depending on the sign of
$m_1-m_2-m_3$. If $m_1-m_2-m_3>0$, the past Milne wedge gives an identical
contribution; the result of adding both is
\eqn\thmodc{ {-i\, \delta_{l_1, l_2+l_3}\over
2 \sqrt{m_1m_2m_3}\; \sqrt{m_1-m_2-m_3}}~.}
This reproduces \thptlim, showing that the non-analyticity is
associated with the asymptotic behavior of the wavefunctions in
the Milne wedges. In the case $m_1-m_2-m_3<0$, the contributions of both Milne
wedges cancel each other, which is consistent with the vanishing of the
three point function in this case.

\appendix{B}{Operator product expansion and infrared divergences}
In this appendix, we compute the OPE of the operators \Nekr, and indicate how
it can be used to explain infrared divergences in four point functions.

We have, using \ppm,
\eqn\opea{\psi_{m^2,l}={1\over 2\sqrt{2} \pi i}\int_{-\infty}^\infty dw
e^{i({m\over\sqrt2}X^-e^{-w}+{m\over\sqrt2}X^+e^{w} +lw)}~.}
The worldsheet scaling dimension of the operator
$\psi_{m^2,l}$ is $(-m^2/4,-m^2/4)$ (we have set $\alpha'=1$). Consider the OPE
\eqn\opeb{\psi_{m_1^2,l_1}(z_1)\psi_{m_2^2,l_2}(z_2)~.}
We can perform the OPE of the plane waves in \opea\ and then integrate
over the $w$'s. The leading term in the OPE is
\eqn\opec{\eqalign{\psi_{m_1^2,l_1}(z_1)\psi_{m_2^2,l_2}(z_2)&=
-{1\over 8\pi^2} \int_{-\infty}^\infty dw_1dw_2|z_1-z_2|^{-{m_1m_2\over2}(
e^{(w_1-w_2)}+e^{(w_2-w_1)})}
e^{i(l_1w_1+l_2w_2)}\times\cr &\times
e^{{i\over\sqrt2}(m_1e^{-w_1}+m_2e^{-w_2})X^-+
{i\over\sqrt2}(m_1e^{w_1}+m_2e^{w_2})X^+} }~.}
We would like to write the r.h.s. as
\eqn\oped{\int d(m^2)
|z_1-z_2|^{(m_1^2+m_2^2-m^2)/2} C(m_1^2,
m_2^2;m^2)\psi_{m^2;l_1+l_2}(z_2)~.}
Comparing \opec\ to \opea, we see that
\eqn\opedd{m^2=(m_1e^{-w_1}+m_2e^{-w_2})
(m_1e^{w_1}+m_2e^{w_2})=m_1^2+m_2^2+
2m_1m_2\cosh(w_1-w_2)~.}
Note that $m^2\ge (m_1+m_2)^2$ for
non-negative $m_1, m_2$ (which we are assuming). The
wavefunction \oped\ has
\eqn\opee{\eqalign{
&p^+={1\over\sqrt2}(m_1e^{-w_1}+m_2e^{-w_2})
\equiv {m\over\sqrt2}e^{-w}~;\cr
&p^-={1\over\sqrt2}(m_1e^{w_1}+m_2e^{w_2}) \equiv
{m\over\sqrt2}e^{w}~.\cr }}
One can write the integral $\int dw_1dw_2$ as $\int d(m^2) dw F(m^2,w)$
where $F$ is the absolute value of the Jacobian of the transformation \opee.
This Jacobian is obtained by using \opee\ to express $w_{1,2}$ as functions of
$m,w$, and then computing the partial derivatives:
\eqn\opeF{F= {1 \over \sqrt{m^4 + m_1^4 + m_2^4 - 2 m^2
m_1^2 - 2 m^2 m_2^2 - 2 m_1^2 m_2^2}}~.}
The square root is precisely the product of the square roots in
the denominator of the three point function~\threepoint, so it
exhibits the same non-analyticity. In appendix A, we argued that
this non-analyticity is an infrared effect. For each $m,w$ there
are two solutions for $w_{1,2}$, and they have to be summed over.
The leading term in the OPE reads
\eqn\opeg{\eqalign{\psi_1(z_1)\psi_2(z_2) & \sim
-{1\over 8\pi^2}\int_{(m_1+m_2)^2}^\infty d(m^2)
|z_1-z_2|^{(m_1^2+m_2^2-m^2)/2} \cr &\times
{1\over\sqrt{m^4+m_1^4+m_2^4-2m^2m_1^2-2m^2m_2^2-2m_1^2m_2^2}}\cr
&\times \Bigl\{\left[{m^2+m_1^2-m_2^2-
\sqrt{m^4+m_1^4+m_2^4-2m^2m_1^2-2m^2m_2^2-2m_1^2m_2^2}\over2mm_1}
\right]^{il_1} \cr &~~\times \left[{m^2-m_1^2+m_2^2+
\sqrt{m^4+m_1^4+m_2^4-2m^2m_1^2-2m^2m_2^2-2m_1^2m_2^2}\over2mm_2}
\right]^{il_2} \cr &~+\left[{m^2+m_1^2-m_2^2+
\sqrt{m^4+m_1^4+m_2^4-2m^2m_1^2-2m^2m_2^2-2m_1^2m_2^2}\over2mm_1}
\right]^{il_1} \cr &~~\times \left[{m^2-m_1^2+m_2^2-
\sqrt{m^4+m_1^4+m_2^4-2m^2m_1^2-2m^2m_2^2-2m_1^2m_2^2}\over2mm_2}
\right]^{il_2} \Bigr\}\cr &\times \int dw e^{i(l_1+l_2)w}
e^{i({m\over\sqrt{2}}X^-e^{-w}
+{m\over\sqrt{2}}X^+e^{w})}~.} }

In what follows, we will need the OPE
\eqn\opebbis{\psi_{m_1^2,l_1}(z_1)\bigl(\psi_{m_2^2,l_2}(z_2)\bigr)^*~.}
This can be easily obtained in the same way as above. The right
hand side of the OPE will now contain all $\psi_{m^2;l_1-l_2}$
with $m^2<(m_1-m_2)^2$. The
Jacobian factor \opeF\ is unchanged; it blows up when
$m^2\approx (m_1-m_2)^2$.

Now consider the four point function \fourptd\ in the kinematical
regime $m_1=m_4,\ m_2=m_3$. It exhibits a divergence from the
region of the integral where $B\approx -A\approx\epsilon$, with
$\epsilon$ small and positive (we will be considering the limit
$\epsilon\to0$). In this region, $v_2, v_3$ and
$v_4$ are all close to 1, and the Jacobian \Jac\ goes like
$1/\epsilon$. This leads to a $\int d\epsilon/\epsilon$ divergence
(for any transverse momenta). In fact, the Jacobian \Jac\ diverges
whenever $v_2=v_3$, which is possible whenever $(m_2-m_3)^2\leq
(m_1-m_4)^2$. However, the Jacobian will only go like $1/\epsilon$
if this inequality is saturated, otherwise it only scales like
$1/\sqrt\epsilon$.

This divergence is an infrared effect, and can be understood using
the OPE we have just described. Assume for simplicity $m_2\geq m_3$, and
consider a process where
particle 2 turns into particle 3, thereby emitting an intermediate
($u$-channel) particle with $m^2=(m_2-m_3)^2-\epsilon$. The OPE
gives a factor $1/\sqrt\epsilon$ from \opeF. The intermediate
particle is now absorbed by particle 1, which turns into particle
4. From the OPE, this process is possible if $(m_2-m_3)^2\leq
(m_1-m_4)^2$. Now there are two possibilities: either there is a
strict inequality, in which case the relevant coefficient in the
second OPE is $\epsilon$-independent; or the inequality is
saturated, in which case the second OPE coefficient also scales
like $1/\sqrt\epsilon$, so that the four point function has a
$\int d\epsilon/\epsilon$ divergence.


\appendix{C}{Comparing the string theory and gravity computations}

In this appendix we complete the computation in section~4 of the
four-point function in dilaton-gravity, and compare it with the
string theory result derived in section~3.
In section~4, we computed the contribution of one of the
four regions of the spacetime $\Rop^{1,1}/\Zop$. In order to obtain the
contribution of the other three regions, we first need to find
useful expressions for the wavefunctions \Nekr\ in those regions.

\subsec{The wavefunctions in the four regions}
We need expressions for the behavior near the singularity of
the wavefunctions \Nekr. For the future Milne wedge, the result is
given in \Nekrleading. This can be extended to the other regions
using the fact that \Nekr\ is built from purely negative frequency
modes in Minkowski space, which are analytic on the lower
half-planes of the complexified horizons. However, in this
appendix we will derive the explicit expressions directly in the
four regions of spacetime.
We label the regions of $\Rop^{1,1}/\Zop$ according to the
sign of $X^+,\,X^-$ as follows:
\item{I)} $X^+,X^-<0$~;
\item{II)} $X^+,X^->0$~;
\item{III)} $X^+>0,X^-<0$~;
\item{IV)} $X^+<0,X^->0$~.

\noindent The wavefunction in all regions is \Nekr\
\eqn\CNekr{
\psi(X^+,X^-)e^{i\vec{p}\cdot \vec{X}}= {1\over 2\sqrt{2} \pi i}\int
dw e^{i(p^+X^-e^{-w}+p^-X^+e^w+lw)}e^{i\vec{p}\cdot\vec{X}}~,}
with $2p^+p^-=m^2$. As before, we will assume that all
the $m$'s and $p$'s are positive, and set $p^+=p^-= m/\sqrt{2}$.

We will use the following integral representations of the Hankel function:
\eqn\CHankint{\eqalign{
H^{(1)}_\nu(z)&={1\over \pi i}e^{-{i\over 2}\pi\nu}\int_{-\infty}^\infty dt\,
e^{ iz\cosh(t)-\nu t},\hskip20pt (z>0)~;\cr
H^{(1)}_\nu(xz)&=-{i\over \pi}e^{-{i\over 2}\nu\pi}z^\nu
\int_0^\infty dt\,
e^{{i\over 2}x(t+z^2/t)}t^{-\nu-1},\hskip20pt  (z=i,\; x>0).\cr}}
We also need the behavior of the wavefunctions $\psi_l$
for small $X^+,X^-$ in all the regions. It is useful to remember the leading
behavior of the Hankel function
$H_{il}^{(1)}(z)$ for small values of the argument $z$:
\eqn\CHank{
H_{il}^{(1)}(z)=J_{il}(z)+iN_{il}(z)\sim
{1\over \sinh(l\pi)}\left[
{e^{l\pi}\over \Gamma(1+il)} \left({z\over 2}\right)^{il}
-{1\over \Gamma(1-il)} \left({z\over 2}\right)^{-il}
\right]~.}
$\underline{{\bf Region\; I:\;} (X^+, X^-)<0}$

\noindent Defining $e^\beta=\sqrt{{p^+X^-\over p^-X^+}}$
we write the wave function as
\eqn\regI{\eqalign{
\psi_l&={1\over 2\sqrt{2} \pi i}\int dw
e^{-i\sqrt{p^+p^-X^+X^-}(e^{-w+\beta}+e^{w-\beta})+ilw}\cr
&={1\over 2\sqrt{2} \pi i} {(p^+X^-/p^-X^+)}^{il/2}
\int dw e^{-i\sqrt{p^+p^-X^+X^-}(e^{-w}+e^w)+ilw}\cr
&={1\over 2\sqrt{2} \pi i}{(p^+X^-/p^-X^+)}^{il/2}
{\biggl(\int dw e^{i\sqrt{p^+p^-X^+X^-}(e^w+e^{-w})-ilw}\biggr)}^*\cr
&={1\over 2\sqrt{2}\pi i} {(p^+X^-/p^-X^+)}^{il/2}
{\biggl( e^{(\pi i/2)(il)}\pi i
H_{il}(2\sqrt{p^+p^-X^+X^-})\biggr)}^*\cr
&={-e^{-\pi l/2}\over 2\sqrt{2}}{(X^-/X^+)}^{il/2}
{\biggl( H_{il}(2\sqrt{p^+p^-X^+X^-}) \biggr)}^*~,\cr}}
where we have used $p^+=p^-$ in the last line.
The behavior near the origin is
\eqn\regIas{
\psi_l\sim-{1\over 2\sqrt{2}}
{1\over\sinh(l\pi)} \left[
 {e^{ \pi l/2} \over \Gamma(1-il)} {\left(-{mX^+\over\sqrt{2}}\right)}^{-il}
-{e^{-\pi l/2} \over \Gamma(1+il)} {\left(-{mX^-\over\sqrt{2}}\right)}^{ il}
\right]~.}
$\underline{{\bf Region\; II:\;} (X^+, X^-)>0}$

\noindent Defining $e^\beta=\sqrt{{p^+X^-\over p^-X^+}}$
we write the wave function here as
\eqn\regII{\eqalign{
\psi_l&={1\over 2\sqrt{2} \pi i}\int dw
e^{i\sqrt{p^+p^-X^+X^-}(e^{-w+\beta}+e^{w-\beta})+ilw}\cr
&={1\over 2\sqrt{2} \pi i} {(p^+X^-/p^-X^+)}^{il/2}\int dw
e^{i\sqrt{p^+p^-X^+X^-}(e^{-w}+e^w)+ilw}\cr
&={1\over 2\sqrt{2}\pi i}{(p^+/p^-)}^{il/2}{(X^-/X^+)}^{il/2} e^{{\pi i\over
2}(-il)}\pi i H_{-il}(2\sqrt{p^+p^-X^+X^-})\cr
&={e^{\pi l\over 2}\over 2\sqrt{2}} {(X^-/X^+)}^{il/2}
H_{-il}(\sqrt{2m^2X^+X^-})~.}}
The leading behavior of the wavefunction near the origin is
\eqn\regIIas{
\psi_l\sim {1\over 2\sqrt{2}}
{-1\over \sinh(\pi l)}
\left[{e^{-\pi l/2} \over \Gamma(1-il)}
{\left({mX^+\over\sqrt{2}}\right)}^{-il} -{e^{\pi l/2} \over \Gamma(1+il)}
{\left({mX^-\over \sqrt{2}}\right)}^{ il} \right]~.}
This reproduces \Nekrleading.

\noindent
$\underline{{\bf Region\; III:\;} (X^+, -X^-)>0}$

\noindent We define $e^\beta=\sqrt{-p^+X^-/p^-X^+}$
and write the leading term of the wave function as
\eqn\regIII{\eqalign{
\psi_l&={1\over 2\sqrt{2}\pi i}
{\left({-p^+X^-\over p^-X^+}\right)}^{il/2}
\int dw e^{i\sqrt{-p^+p^-X^+X^-}(-e^{-w+\beta}+e^{w-\beta})+il(w-\beta)}\cr
&\sim{1\over 2\sqrt{2}\pi i}
{\left({-X^- \over X^+}\right)}^{il/2} (\pi i)
e^{\pi l/2} {(i)}^{il}{-1\over \sinh(l\pi)}\cr
&\times\left[
{e^{-l\pi}\over \Gamma(1-il) }
\left({im\sqrt{X^+(-X^-)}\over\sqrt{2}}\right)^{-il}-
{        1\over \Gamma(1+il) }
\left({im\sqrt{X^+(-X^-)}\over\sqrt{2}}\right)^{ il}\right]\cr
&={1\over 2\sqrt{2}}
{-1\over \sinh(l\pi)}
\left[ {e^{ -l\pi /2}\over \Gamma(1-il)}
{\left({m\over\sqrt{2}}\right)}^{-il} (X^+)^{-il} -
{e^{-l\pi /2}\over \Gamma(1+il)}
{\left({m\over\sqrt{2}}\right)}^{ il}
(-X^-)^{ il} \right]~.\cr}}
$\underline{{\bf Region\, IV:}\, (-X^+, X^-)>0}$

\noindent
The behavior of the wave function near the origin is:
\eqn\regIV{
\psi_l\sim -{1\over 2\sqrt{2}}
{1\over \sinh(\pi l)}
\left[ {e^{\pi l/2}\over \Gamma(1-il)}
{\left({m\over\sqrt{2}}\right)}^{-il}
{(-X^+)}^{-il}
-{e^{\pi l/2}\over \Gamma(1+il)} {\left({m\over\sqrt{2}}\right)}^{il}
{(X^-)}^{il}\right]~.}
\subsec{The four point function}
We are now ready to compute the four-point function in gravity and compare it
to the string result. The four-point function, at least in the kinematic regime of
section~4, is~\fourptwxbis\
\eqn\Cfourptwx{\eqalign{
\langle\psi^*_3\psi^*_4\psi_1\psi_2\rangle&=\cr
&-(2\pi)^{24}\delta^{24}(\sum \epsilon_i\vec p_i)2\pi\int dX^+dX^-
\partial_-\psi_{m_1,l_1}\partial_-\psi^*_{m_3,l_3} {1\over\partial^2}
\partial_+\psi_{m_2,l_2}\partial_+\psi_{m_4,l_4}^*~,
}
}
where
\eqn\Coneoverbox{
{1\over\partial^2}=-{1\over(\vec p_2-\vec p_4)^2}~.
}
The contribution to the four-point function from the region near the
singularity may be evaluated by simply inserting the expressions for the
leading behavior of the wavefunction from the various
regions~\regIas-\regIV.
Define
\eqn\CSdef{
S(m_i,l_i,p_i) =
{(2\pi)^{24}\delta^{24}(\sum \epsilon_i\vec p_i) 2\pi
l_1l_2l_3l_4
\Bigl({m_1\over\sqrt2}\Bigr)^{il_1}
\Bigl({m_2\over\sqrt2}\Bigr)^{-il_2}
\Bigl({m_3\over\sqrt2}\Bigr)^{-il_3}
\Bigl({m_4\over\sqrt2}\Bigr)^{il_4}
\over 64 (\prod\sinh(\pi l_i))
\Gamma(1+il_1)\Gamma(1-il_2)\Gamma(1-il_3)\Gamma(1+il_4)
(\vec p_1-\vec p_3)^2},
}
in terms of which the contribution to the four-point function from the
various regions are

\noindent $\underline{{\bf Region\; I:}}$
\eqn\regIterm{
S(m_i,l_i,p_i)e^{\pi (-l_1+l_2-l_3+l_4)/2}
\int_{-\infty}^0 dX^+\int_{-\infty}^0 dX^-
{(-X^+)}^{i(l_4-l_2)-2}{(-X^-)}^{i(l_1-l_3)-2}~;}
$\underline{{\bf Region\; II:}}$
\eqn\regIIterm{
S(m_i,l_i,p_i)e^{\pi (l_1-l_2+l_3-l_4)/2}
\int_0^{\infty} dX^+\int_0^{\infty} dX^-
{(X^+)}^{i(l_4-l_2)-2}{(X^-)}^{i(l_1-l_3)-2}~;}
$\underline{{\bf Region\; III:}}$
\eqn\regIIIterm{
S(m_i,l_i,p_i)e^{\pi (-l_1-l_2-l_3-l_4)/2}
\int_0^{\infty} dX^+\int_{-\infty}^0 dX^-
{(X^+)}^{i(l_4-l_2)-2}{(-X^-)}^{i(l_1-l_3)-2}~;}
$\underline{{\bf Region\; IV:}}$
\eqn\regIVterm{
S(m_i,l_i,p_i)e^{\pi (l_1+l_2+l_3+l_4)/2}
\int_{-\infty}^0 dX^+\int_0^{\infty} dX^-
{(-X^+)}^{i(l_4-l_2)-2}{(X^-)}^{i(l_1-l_3)-2}~.}
The complete four-point function is therefore
\eqn\Cfourptfull{\eqalign{
\langle\psi^*_3\psi^*_4\psi_1\psi_2\rangle&=
{(2\pi)^{24}\delta^{24}(\sum \epsilon_i\vec p_i) 2\pi l_1l_2l_3l_4
\Bigl({m_1\over\sqrt2}\Bigr)^{il_1}
\Bigl({m_2\over\sqrt2}\Bigr)^{-il_2}
\Bigl({m_3\over\sqrt2}\Bigr)^{-il_3}
\Bigl({m_4\over\sqrt2}\Bigr)^{il_4}
\over 64 (\prod\sinh(\pi l_i))
\Gamma(1+il_1)\Gamma(1-il_2)\Gamma(1-il_3)\Gamma(1+il_4)
(\vec p_1-\vec p_3)^2}\cr
&\times\left(e^{\pi (-l_1+l_2-l_3+l_4)/2} + e^{\pi (l_1-l_2+l_3-l_4)/2}
+ e^{\pi (-l_1-l_2-l_3-l_4)/2} + e^{\pi (l_1+l_2+l_3+l_4)/2}\right)\cr
&\times\int_0^{\infty} dX^+\int_0^{\infty} dX^-
{(X^+)}^{i(l_4-l_2)-2}{(X^-)}^{i(l_1-l_3)-2}~.}}
Using $X^{\pm} = {1\over\sqrt{2}}e^{\eta\pm x}$, we can write the integral
in the above expression as
\eqn\Cnewint{\eqalign{
\int_0^{\infty} dX^+\int_0^{\infty} dX^-
&{(X^+)}^{i(l_4-l_2)-2}{(X^-)}^{i(l_1-l_3)-2} =\cr
&2 \int_{-\infty}^{\infty} d\eta\int_{-\infty}^{\infty} dx
\left(e^{\eta}\over\sqrt{2}\right)^{i(l_1+l_4-l_2-l_3)-2}
e^{ix(l_3+l_4-l_1-l_2)}~.}}
Doing the $x$ integral, and defining $v=2e^{-2\eta}$, we can simplify this
to
\eqn\Cnewinta{
2\pi\delta(l_1+l_2-l_3-l_4)\int_0^{\infty}dv\, v^{i(l_2+l_3-l_1-l_4)/2}~.}
Finally, writing $v= \half m_1m_4 u$, the four point function becomes
\eqn\Cfourptnew{\eqalign{
&\langle\psi^*_3\psi^*_4\psi_1\psi_2\rangle=
{(2\pi)^{24}\over 16}\delta^{24}(\sum \epsilon_i\vec p_i) \delta(\sum \epsilon_il_i)
{m_1m_4 \over (\vec p_1-\vec p_3)^2}
\Bigl({m_4\over m_2}\Bigr)^{il_2}
\Bigl({m_3\over m_1}\Bigr)^{-il_3}
\int_0^{\infty}du\, u^{i(l_2-l_4)}\cr
&\times{\pi^2l_1l_2l_3l_4(e^{\pi (-l_1+l_2-l_3+l_4)/2} + e^{\pi (l_1-l_2+l_3-l_4)/2}
+ e^{\pi (-l_1-l_2-l_3-l_4)/2} + e^{\pi (l_1+l_2+l_3+l_4)/2})\over
2 (\prod\sinh(\pi l_i))
\Gamma(1+il_1)\Gamma(1-il_2)\Gamma(1-il_3)\Gamma(1+il_4)}.
}}
This expression is identical to the $\alpha'\to 0$ limit of
the string theory result~\fourptf\ provided we rescale
\eqn\uvfour{
u = A(l_i)v_4~,}
where the coefficient $A(l_i)$ is independent of the masses $m_i$.

As a consistency check, this rescaling does nothing in the
(unphysical) case $1+i(l_2-l_4) = 1+i(l_3-l_1) = 0$. Therefore, we
expect that for these values of $l_i$, the second line of
\Cfourptnew\ should not depend on $l_i$. In fact, using standard
Gamma function identities, it is easy to show that
\eqn\Ccoeff{\eqalign{ &{\pi^2l_1l_2l_3l_4(e^{\pi
(-l_1+l_2-l_3+l_4)/2} + e^{\pi (l_1-l_2+l_3-l_4)/2} + e^{\pi
(-l_1-l_2-l_3-l_4)/2} + e^{\pi (l_1+l_2+l_3+l_4)/2})\over 2
(\prod\sinh(\pi l_i))
\Gamma(1+il_1)\Gamma(1-il_2)\Gamma(1-il_3)\Gamma(1+il_4)}\cr
&={\pi^2(-i+l_3)l_2l_3(-i+l_2) \left(e^{\pi(l_1-l_4)} +
e^{-\pi(l_1-l_4)} - e^{-\pi(l_1+l_4)} -e^{\pi(l_1+l_4)}\right)
\over 2 \Pi_i\, \sinh(\pi l_i)
\Gamma(2+il_3)\Gamma(1-il_3)\Gamma(1-il_2)\Gamma(2+il_2)}\cr
&=-{\pi^2(1+il_3)l_2l_3(1+il_2) \left(e^{\pi(l_1-l_4)} +
e^{-\pi(l_1-l_4)} - e^{-\pi(l_1+l_4)} -e^{\pi(l_1+l_4)}\right)
\over 2\Pi_i\, \sinh(\pi l_i)
\Gamma(2+il_3)\Gamma(1-il_3)\Gamma(1-il_2)\Gamma(2+il_2)} \cr
&=-{\pi^2l_2l_3 \left(e^{\pi(l_1-l_4)} + e^{-\pi(l_1-l_4)} -
e^{-\pi(l_1+l_4)} -e^{\pi(l_1+l_4)}\right) \over 2 \Pi_i\,
\sinh(\pi l_i)
\Gamma(1+il_3)\Gamma(1-il_3)\Gamma(1-il_2)\Gamma(1+il_2)} \cr
&={e^{\pi(l_1+l_4)} + e^{-\pi(l_1+l_4)} - e^{\pi(l_1-l_4)}
-e^{-\pi(l_1-l_4)}\over 2 \sinh(\pi l_4)\sinh (\pi l_1)}\cr
&=2~.}} \listrefs
\end